\documentclass[prb,reprint,aps,superscriptaddress,longbibliography]{revtex4-2}
\usepackage[colorlinks=true,linkcolor=blue,citecolor=blue,urlcolor=blue]{hyperref}
\usepackage{graphicx}
\usepackage{bm}
\usepackage{braket}
\usepackage{amsmath}
\usepackage{amssymb}
\usepackage{graphics}
\usepackage{enumerate}
\usepackage{siunitx}
\usepackage[dvipsnames]{xcolor}
\usepackage{extarrows}

\begin{document}
\title{Lifetime and spectral function of topological heavy fermions}
\author{Nemin Wei}
\affiliation{\mbox{Department of Physics and Yale Quantum Institute, Yale University, New Haven, Connecticut 06520, USA}}

\author{Felix von Oppen}
\affiliation{\mbox{Dahlem Center for Complex Quantum Systems, Fachbereich Physik, and Halle-Berlin-Regensburg}\\ Cluster of Excellence CCE, Freie Universit\"at Berlin, 14195 Berlin, Germany}

\author{Leonid I.\ Glazman}
\affiliation{\mbox{Department of Physics and Yale Quantum Institute, Yale University, New Haven, Connecticut 06520, USA}}

\date{\today}

\begin{abstract}
Twisted bilayer graphene provides a paradigmatic platform for exploring the interplay between electronic topology and strong correlations. Within the topological heavy fermion model [Song and Bernevig, Phys. Rev. Lett. 129, 047601 (2022)], topology and electron interactions are brought together by including a weak hybridization between the bands of itinerant $c$- and localized $f$-electrons. Hybridization infuses concentrated Berry curvature into the $f$-band, while leaving it flat. These band features have motivated recent proposals of a Mott semimetal phase above the flavor-ordering temperature at charge neutrality.
In this work, we develop an analytic theory of the quasiparticle dispersion and lifetime in the Mott semimetal. We reformulate the interacting flat-band Hamiltonian as an on-site Hubbard interaction defined on a set of non-orthogonal orbitals, and compute the electron Green’s function using the equation-of-motion method, in close analogy with the Hubbard-III approximation. Unlike the conventional Hubbard model, in our case this approximation is controlled by a well-defined small parameter in the theory. We evaluate the electron self-energy and demonstrate the emergence of well-defined low-energy quasiparticles with the dispersion and relaxation rate proportional to the interaction strength. The quasiparticle spectrum is well-resolved in energy and in momentum down to the very vicinity of the Fermi level.
Our results illustrate unconventional spectral properties arising from strong correlations and nontrivial quantum geometry, and have direct relevance for spectroscopic probes such as quantum twisting microscope experiments.

\end{abstract}

\maketitle

\section{Introduction}
The prediction \cite{bistritzer2011moire} and experimental discovery \cite{cao2018correlated,cao2018unconventional} of flat electronic bands in magic-angle twisted bilayer graphene (MATBG) introduced a highly versatile platform for studying electron correlations. A large number of interaction-driven effects were observed in quick succession \cite{lu2019superconductors,cao2021nematicity,oh2021evidence,jaoui2022quantum}. These include symmetry-breaking Stoner-type instabilities, emblematic for itinerant electrons, which result in valley- and spin-polarized states, see, e.g., Ref.\  \cite{sharpe2019emergent}. Some of the valley-polarized states give rise to the (quantized) anomalous Hall effect~\cite{sharpe2019emergent,serlin2020intrinsic,stepanov2021competing,lin2022spin,xie2021fractional}, attesting to the nontrivial topology of the moir\'e electron band structure \cite{bultinck2020ground,po2018origin,zou2018band,song2021twisted}. Other phenomena were associated with the interaction-induced electron localization, including observations of the Pomeranchuk effect \cite{saito2021isospin,rozen2021entropic} as well as the Mott transition \cite{wong2020cascade,zondiner2020cascade}. 

The topological nature of the flat bands appears to be in tension with the phenomenology of localized electrons. One approach to resolving this tension is to construct exponentially localized orbitals by incorporating both flat and remote bands \cite{haule2019the,datta2023heavy,carr2019derivation}, as in the topological heavy-fermion model \cite{song2022magic}. Alternatively, one may work directly with nonlocal Wannier orbitals of the flat bands, which exhibit power-law-decaying but parametrically small tails \cite{zang2022real,ledwidth2025nonlocal}. Reference \cite{ledwidth2025nonlocal} shows that these Wannier orbitals can host nonlocal moments driven by strong on-site repulsion.
The results of the two approaches are consistent with each other in predicting a remarkable reconstruction of the elementary excitation spectrum, even without an intervening symmetry-breaking transition. 


A striking prediction from both numerical \cite{hofmann2022fermionic,haule2019the,datta2023heavy,rai2024dynamical} and analytical \cite{ledwidth2025nonlocal,hu2025projected,zhao2025topological} studies is the migration of the low-energy spectral weight 
from the Dirac points to the $\Gamma$ point in the moir\'e Brillouin zone (mBZ) of MATBG at charge neutrality. In the flavor unpolarized Mott semimetal phase \cite{ledwidth2025nonlocal}, the quasiparticle spectrum reaches zero at the $\Gamma$ point. Sufficiently far from the $\Gamma$ point, at $k\gtrsim k_\star$, the quasiparticle dispersion becomes flat with the excitation energy proportional to the strength of the electron-electron interaction, $\epsilon_{ k}\to\pm U/2$. The value of $k_\star$  characterizes the concentration of Berry curvature around the $\Gamma$ point. In the THF model \cite{hu2025projected}, $k_\star\sim\gamma/v$ characterizes the region of strong hybridization between the flat zero-energy and the dispersive bands. ($\gamma$ is the hybridization energy and $v$ is the velocity in the dispersive bands).

The identification of elementary excitations of a strongly-interacting electron system inevitably raises the question of their lifetimes and their affinity to the free-electron excitations. At a qualitative level, this question was partially addressed in Ref.~\cite{ledwidth2025nonlocal}. This work introduces a phenomenological parameter $s\sim a_M k_\star$ in terms of the lattice constant $a_M$ of the moir\'e pattern, and conjectured an energy broadening of order $s^2U$ assuming $s\ll 1$. Recently, this estimate was refined with the result that the spectral width for states in the flat ``high-energy'' part ($k\gtrsim k_\star$) of the spectrum is of the order of $Ns^2U$, where $N=8$ is the number of quasiparticle flavors in an unpolarized state \cite{vituri2026aps}. 
However, the intriguing question of the ``quality'' of the low-energy elementary excitations close to the $\Gamma$ point of the mBZ remained open. The goal of our work is to evaluate the lifetime of excitations in that part of the spectrum.

We base our work on the THF model~\footnote{We are aware of two other works \cite{vituri2026controlled,hu2026thf} addressing a similar set of questions by different techniques.}. 
The model is controlled by a small number of parameters which have clear physical interpretations. In a minimal setting, these include the already mentioned velocity $v$ of the Dirac spectrum  of the dispersive bands, their hybridization energy $\gamma$ with the localized $f$ orbitals, and the on-site interaction energy $U$  of the $f$-electrons. In addition, weak hybridization $M$ between the two dispersive Dirac bands in each spin-valley sector 
introduces a nonzero single-particle bandwidth of the MATBG flat bands.
A recent measurement of the MATBG band structure using the quantum twisting microscope  \cite{xiao2025interacting} suggests $M\ll U/2\lesssim\gamma$. The phenomenology of Ref.~\cite{ledwidth2025nonlocal} was developed under the stronger assumption $M\ll U\ll\gamma$. We use this condition to project the full THF Hamiltonian onto the flat bands separated by a gap $\sim \gamma$ from the rest of the spectrum. 

The structure of the projected Hamiltonian resembles that of a Hubbard model in the limit of strong repulsion. Like in the latter, the dominant term in the projected Hamiltonian is the onsite repulsion $U$. Unlike the conventional Hubbard model with short-range single-particle hopping, we encounter a weak, but long-range many-body hopping term. In the projected THF model, electron hopping occurs due to the presence of a small nonlocal two-particle interaction. It occurs with amplitude $\sim s^2U$ and can be accompanied by a change of the state of another electron. Remarkably, this generalized Hubbard model hosts well-defined quasiparticle states and allows for a perturbative treatment, assuming the parameter $Ns^2$ is small, $Ns^2\ll 1$.

At $M=0$, all terms of the projected THF Hamiltonian are proportional to the interaction $U$. At $U=0$ the states are heavily degenerate (or almost-degenerate at a finite $M\ll U$). This makes application of diagrammatic theory difficult. Instead, we employ equations of motion for the Green's functions as in Hubbard's original work \cite{hubbard1963I,hubbard1964III}. These equations form a chain relating the time evolution of a lower-order correlation function to a higher-order one. 


Truncating this chain after the second equation allows one to find the electron self-energy $\Sigma (k, \omega)$ in what is known as the Hubbard-I approximation \cite{hubbard1963I}. At the Hubbard-I level, $\Sigma (k, \omega)$ is a real-valued function along the real frequency axis, giving rise to an unbroadened
quasiparticle spectrum, defined to the zeroth order in $s$. Our results for the quasiparticle dispersion relation agree with the spectra found in Ref.~\cite{ledwidth2025nonlocal,hu2025projected} in the absence of symmetry breaking. Specifically, at $M=0$ and at $k\ll k_\star$ the dispersion is linear, $\epsilon_{k}= \pm (U/2)k/k_\star$. In the opposite limit, $k\gg k_\star$, it approaches a constant, $\epsilon_{k}\approx \pm U/2$. The presence of $M\neq 0$ modifies the dispersion from linear to quadratic at the lowest energies, $\epsilon_{k}\lesssim M$, in a narrow vicinity, $k\lesssim (M/U)k_\star$, of the $\Gamma$ point.

To find our key results for the quasiparticle lifetime, we keep two more consecutive levels of the chain of equations. Factorizing the highest-order correlation function into lower-order ones allows us to obtain a closed system of equations which correspond to the Hubbard-III approximation. This approximation generates self-energy corrections of order $s^2$ relative to the Hubbard-I result. Importantly, $\Sigma(k, \omega)$ acquires an imaginary part, leading to a finite quasiparticle lifetime. We find
the dependence of the relaxation rate $1/\tau_k$ on the quasiparticle momentum for the entire band. Our most important finding is that the linear part of the dispersion relation remains well-resolved, $\epsilon_{k}\tau_{k}\sim (Ns^2)^{-1}\gg 1$, leading to a sharp peak in the spectral function down to energies $\epsilon_{k}\sim M$. At lower energies, the peak broadens and becomes weaker indicating a qualitative difference between the quasiparicle and free-electron states.

Below, we first present our main results without entering technical details in Sec.\ II. The equation of motion methods and Hubbard-I approximation are introduced  in Sec.\ III. Section IV provides details of the theory at the Hubbard III level for the projected THF Hamiltonian with $M=0$. The extension of the theory to  $M\neq 0$ is described in Sec.\ V. We conclude in Sec.\ VI.

\section{Main Results}
%
%
The THF model consists of a periodic lattice of localized $f$-electrons hybridizing with itinerant $c$-electrons. We use $f_{i \sigma}$ to denote the annihilation operators of localized $f$-electrons on site $\bm R_i$. The flavor index $\sigma=(\eta\alpha)$ combines $\eta$ specifying spin and valley with the orbital index $\alpha=1,2$. The Hamiltonian of the THF model, $H = H_0 + H_U$, consists of the free-particle term \cite{hu2025projected},
\begin{widetext}
    \begin{align}\label{eq:H_0_full}
    H_0 = \sum_{\eta}\sum_{\bm k} (f_{\bm k \eta}^{\dagger}, c_{\bm k 1\eta }^{\dagger}, c_{\bm k 2\eta }^{\dagger})
    \begin{pmatrix}
        0 & \gamma\rho_0 &0\\
        \gamma\rho_0 & 0 & v(k_x\rho_0 + i k_y\rho_z)\\
        0 & v(k_x\rho_0-ik_y\rho_z)& M\rho_x\\
    \end{pmatrix}
    \begin{pmatrix}
        f_{\bm k \eta }\\ 
        c_{\bm k 1\eta }\\ 
        c_{\bm k 2\eta }
    \end{pmatrix},
\end{align}
and the interaction
\end{widetext}
\begin{align}\label{eq:localU}
    H_{U}=\frac{U}{2} \sum_{i}\left(\sum_{\sigma=1}^{N}f_{i \sigma}^{\dagger}f_{i \sigma} - \frac{N}{2}\right)^2.
\end{align}
Here $\rho_i$ denotes Pauli matrices in orbital space, and $$f_{\bm k \eta\alpha}= \frac{1}{\sqrt{N_s}}\sum_{i}e^{-i\bm k\cdot \bm R_i}f_{i \eta \alpha}$$
annihilates $f$-electrons at quasimomentum $\bm k$, with $N_s$ being the number of sites.  
Compared to the full THF model of TBG \cite{song2022magic},  we have made a few simplifications: (i) we use a momentum-independent hybridization strength $\gamma$ between the $c$-and $f$-electrons; (ii) we neglect intersite repulsion between $f$-electrons, and treat the $c$-electrons as noninteracting.

In the chiral-flat limit, $M=0$, the orbitals $\alpha=1,2$ decouple at the single-particle level. Within each spin, valley, and orbital sector, $H_0$ has a zero-energy band and a pair of remote conduction ($l=+$) and valence ($l=-$) bands at energies 
\begin{equation}
 \epsilon_{\bm k}^{l=\pm}=\pm v\sqrt{k^2+k_{\star}^2}   
\end{equation}
with $k=|\bm k|$ and $k_{\star}=\gamma/v$. The annihilation operators of the zero-energy and the remote bands are given by
\begin{equation}\label{def:d}
        d_{\bm k\eta\alpha} = \frac{k}{\sqrt{k^2+k_{\star}^2}}f_{\bm k\eta\alpha} - \frac{[k_x- i(-)^{\alpha} k_y]k_{\star}}{k\sqrt{k^2+k_{\star}^2}}c_{\bm k2\eta\alpha},
\end{equation}
%
%
%
and
\begin{equation}
\begin{split}
    a_{\bm k \eta\alpha l } = \frac{1}{\sqrt{2}}\Big[&\frac{k_{\star}}{\sqrt{k^2+k_{\star}^2}}f_{\bm k\eta\alpha }\\
    &+ l c_{\bm k1\eta\alpha}+\frac{k_x-i(-)^{\alpha}k_y}{\sqrt{k^2+k_\star^2}}c_{\bm k2\eta\alpha}\Big],
\end{split}
\end{equation}
%
respectively. These operators approximately respect the periodicity of the mBZ, provided that the dimensionless parameter 
\begin{equation}
    s^2 = \frac{\pi k_{\star}^2}{\Omega_m}
\end{equation}
is small, $s^2\ll 1$. Here $\Omega_m$ denotes the mBZ area.

We now treat $M$ and $U$ as perturbations, which are small compared to the band gap $\gamma$, $\gamma\gg M,U>0$. We can then project $H$ into the nearly-zero-energy bands [cf., Eq.~\eqref{def:d}], while keeping the remote valence (conduction) band fully occupied (empty). This is effected by a projection operator $P$, which obeys
\begin{equation}
    P d_{\bm k \sigma} = d_{\bm k \sigma}P,\quad Pa_{\bm k  \sigma- } =a_{\bm k \sigma+}P =0.
\end{equation}
%
The projected single-particle Hamiltonian reads
\begin{align}
    &P H_0P =\sum_{\sigma,\sigma'} \sum_{\bm k}d_{\bm k \sigma}^{\dagger}h(\bm k)_{\sigma,\sigma'}d_{\bm k \sigma'},\label{eq:PH0P}\\
    &h(\bm k)_{\eta\alpha,\eta'\alpha'}= \frac{Mk_{\star}^2}{k^2+k_{\star}^2}\delta_{\eta,\eta'}
    \begin{pmatrix}
        0 & e^{-2i\theta_{\bm k}}\\
        e^{2i\theta_{\bm k}}&0\\
    \end{pmatrix}_{\alpha,\alpha'},\label{eq:h}
\end{align}
with $e^{\pm i\theta_{\bm k}}=(k_x\pm ik_y)/k$. We note that at $M=0$, the bands remain perfectly flat.

The projected on-site interaction $\tilde{H}_U=PH_UP$ reads
%
\begin{align}\label{eq:PHP}
    \tilde{H}_U=&\frac{U}{2}\sum_{\{\bm k_i\}}\sum_{\sigma,\sigma'}\delta_{\lfloor{\bm k_1+ \bm k_2}\rfloor,\lfloor{\bm k_3+\bm k_4}\rfloor}\left(\prod_{i=1}^4\frac{k_i}{\sqrt{k_i^2+k_{\star}^2}}\right)\notag\\
    &\quad\times(d_{\bm k_1\sigma}^{\dagger}d_{\bm k_4\sigma}-\frac{\delta_{\bm k_1,\bm k_4}}{2})(d_{\bm k_2\sigma'}^{\dagger}d_{\bm k_3\sigma'}-\frac{\delta_{\bm k_2,\bm k_3}}{2}),
\end{align}
where $\lfloor\bm k\rfloor$ denotes the momentum equivalent to $\bm k$ in the first mBZ.
The nontrivial quantum geometry of the flat bands results in momentum-dependent form factors: the interaction strength becomes almost constant for $k_{i}\gg k_{\star} \ (i=1,2,3,4)$, but vanishes if any of the four momenta approaches zero. This reflects that $d_{\bm k=0}$ consists entirely of $c$-electrons and is unaffected by the interaction $H_U$ between $f$-electrons.

While we started with an on-site interaction Eq.~(\ref{eq:localU}) of $f$-electrons, the band hybridization built into the projected Hamiltonian, Eqs.~(\ref{eq:h}) and (\ref{eq:PHP}), makes the interaction of the projected model nonlocal. The nonlocality is characterized by the function
\begin{equation}\label{eq:lambda}
    \lambda_{ij} = \frac{1}{z}\int_{\text{mBZ}}\frac{d^2k}{\Omega_m} \frac{k^2}{k^2+k_{\star}^2}e^{i\bm k\cdot(\bm R_i-\bm R_{j})},
\end{equation}
where the normalization constant $z$ [see Eq.~\eqref{eq:z} below] ensures $\lambda_{ii}=1$. We note that $\lambda_{ij}$ decays exponentially with $|\bm R_i-\bm R_j|$, but over a large characteristic length scale $a_M/s$. Below this length scale, $\lambda_{ij}  \sim s^2$ for $i\neq j$. The nonlocal two-particle interaction facilitates electron hopping. This leads to a dispersive quasiparticle band even at $M=0$, where single-particle hopping is fully absent. Interaction-assisted hopping of one electron depends on the state of another electron, which ultimately results in a finite quasiparticle lifetime.

%
\begin{figure}
    \centering
    \includegraphics[width=1\linewidth]{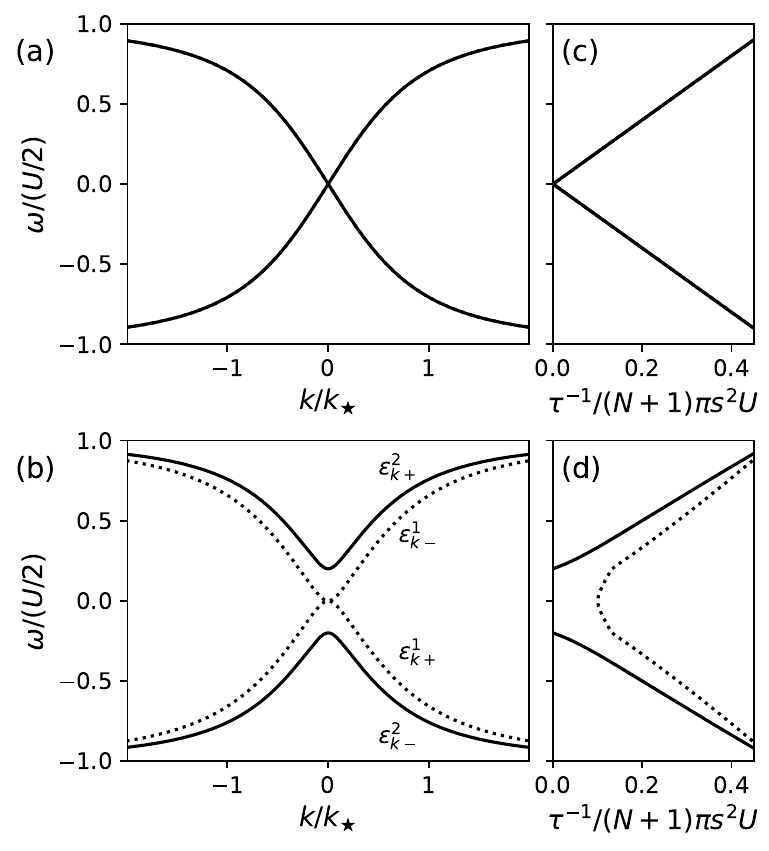}
    \caption{Quasiparticle dispersion and relaxation rate of the THF model at charge neutrality. (a) Dispersion in the chiral-flat limit, $M=0$,  obtained within the Hubbard-I approximation [Eq.~\eqref{eq:twomodes}]; (b) Dispersion at $M=0.1 U$. The solid (dashed) curves plot the dispersion $\epsilon_{\bm k \pm}^{2}$ ($\epsilon_{\bm k \pm}^{1}$) [Eqs.~\eqref{eq:fourmodes}], whose quasiparticle residue approaches $1$ $(0)$ as $\bm k\rightarrow 0$. (c), (d) Relaxation rate $1/\tau_{\bm {k}\pm}^{2}$ (solid) and $1/\tau_{\bm{k} \pm}^{1}$ (dashed) as a function of quasiparticle energy at $M=0$ (c) and $M=0.1 U$ (d) [Eqs.~\eqref{eq:rateMzero} and~\eqref{eq:rate_M}, respectively].}
    \label{fig:lifetime}
\end{figure}

The Hamiltonian in Eqs.~(\ref{eq:h})-(\ref{eq:PHP}) is reminiscent of a Hubbard model with strong on-site repulsion ($U/t\gg$ 1), except that the nature of hopping is different. In the original Hubbard model, the matrix element $t$ controls the short-range one-electron hopping. In contrast, in our case electron hopping occurs solely (at $M=0$) or predominantly (at $M\ll U$) due to the interaction, via its weaker, nonlocal component. The similarity to the Hubbard model motivates us to apply the method of equations of motion used in Hubbard's original  work~\cite{hubbard1963I, hubbard1964III}. We use this method to find the Green's function $\langle\langle d_{\bm k \sigma}; d_{\bm k \sigma}^{\dagger}\rangle\rangle_{\omega}$ in the projected THF model at half-filling of the $N$ flat bands, i.e., at charge neutrality. In evaluating the electron self-energy, we assume a flavor-symmetric, i.e., disordered, state -- similar to the spin-disordered state of the Hubbard model with a $U/t\gg 1$ at half filling.

Evaluation of the self-energy $\Sigma (\bm k,\omega)$ at the Hubbard-I level allows us to find the quasiparticle dispersion, 
\begin{equation}\label{eq:twomodes}
\epsilon_{\bm k} = \pm \frac{U}{2}\frac{k}{\sqrt{k^2+k_{\star}^2}},
\end{equation}
at $M=0$, see Fig.~\ref{fig:lifetime}(a). The respective poles of $\langle\langle d_{\bm k \sigma}; d_{\bm k \sigma}^{\dagger}\rangle\rangle_{\omega}$ lie on the real-$\omega$ axis and have a momentum-independent residue $1/2$, similar to the conventional Hubbard model. 
At that level, the self-energy $\Sigma_{\rm I}(\bm k, \omega)$ is not affected by $M\neq 0$, but its presence results in splitting of the degeneracy which remained in each of the modes~(\ref{eq:twomodes}).  Now the Green's function has four branches of poles at energies
\begin{subequations}\label{eq:fourmodes}
    \begin{align}
        \epsilon_{\bm k\pm}^{1} = \pm\frac{|\hat{t}_{\bm k}|}{2}&  \mp \sqrt{\frac{|\hat{t}_{\bm k}|^2}{4}+\frac{U^2}{4}\frac{k^2}{k^2+k_{\star}^2}},\label{eq:epsilon_1}\\
        \epsilon_{\bm k\pm}^{2} = \pm\frac{|\hat{t}_{\bm k}|}{2}& \pm \sqrt{\frac{|\hat{t}_{\bm k}|^2}{4}+\frac{U^2}{4}\frac{k^2}{k^2+k_{\star}^2}},\label{eq:epsilon_2}
    \end{align}
\end{subequations}
where
\begin{equation}
            \hat{t}_{\bm k} = M \frac{k_{\star}^2}{k^2+k_{\star}^2}e^{-2i\theta_{\bm k}}.
\end{equation}
The corresponding quasiparticle residues take the form
\begin{equation}
    Z_{\bm k\pm}^{i} =\frac{|\epsilon_{\bm k\pm}^{i}|}{\sqrt{M^2\big[\frac{k_{\star}^2}{k^2+k_{\star}^2}\big]^2+U^2\frac{k^2}{k^2+k_{\star}^2}}},\ (i=1, 2).
\end{equation}
Figure~\ref{fig:lifetime}(b) depicts the four branches of the dispersion. Modes $\epsilon_{\bm k\pm}^{2}$ are direct descendants of the free-particle excitations, as in the limit $U\to 0$ they become eigenvalues of the Hamiltonian Eq.~(\ref{eq:h}). In contrast, the modes $\epsilon_{\bm k\pm}^{1}$ emerge as a result of a finite-strength interaction $U\neq 0$. We find that despite the single-particle splitting $\pm M$ at $\bm k=0$, the charge excitation spectra $\epsilon_{\bm k\pm}^{1}$ of the paramagnetic Mott state remain gapless at the origin, $\epsilon_{\bm k\pm}^{1}\sim \mp U^2k^2/4Mk_{\star}^2$.  However, the respective quasiparticle residue is rapidly suppressed,  $Z_{\bm k\pm}^{1}\ll 1$, as $k\ll (M/U)k_{\star}$ approaches the $\Gamma$ point.
This agrees with the notion 
that these low-energy excitations are predominantly of a trion character \cite{ledwith2025exotic}. The dispersion relations in  Eqs.~(\ref{eq:twomodes}) and (\ref{eq:fourmodes}) agree with the Mott-semimetal spectra obtained in the phenomenological description of Ref.\ \cite{ledwidth2025nonlocal,hu2025projected}.

The main advance of our work is the evaluation of the next-order, Hubbard-III approximation to the electron self-energy. In the original Hubbard model, this approximation is not controlled, no matter how small the ratio $t/U$. Remarkably, in our case the approximation {\sl is} controlled by the small parameter $s^2$.  To the first order in $s^2$, the self-energy $\Sigma_{\rm III}(\bm k, \omega)$ acquires an imaginary part at the quasiparticle spectrum, $\omega=\epsilon_{\bm k}$, for each of the branches in Eqs.\ \eqref{eq:twomodes} and \eqref{eq:fourmodes}. We provide the full expressions for $\Sigma_{\rm III}(\bm k, \omega)$ in Eqs.~(\ref{eq:self-energy_HubbardIII}) and~(\ref{eq:self-energy_HubbardIIIM}). Here, we focus on the results for the relaxation rates as obtained from these equations. 

At $M=0$, the quasiparticle relaxation rate is
\begin{equation}\label{eq:rateMzero}
    \frac{1}{\tau_{\bm k}}= (N+1)\pi s^2|\epsilon_{\bm k}|.
\end{equation}
The relaxation rate is proportional to the energy but parametrically smaller. This indicates that the energy of the quasiparticle pole is well-resolved at all momenta. The momentum resolution of the dispersion relation is controlled by the dimensionless parameter $(kl_{\bm k})^{-1}$ involvong the quasiparticle mean free path $l_{\bm k}$. We find
\begin{equation}
    \frac{1}{kl_{\bm k}} = \frac{|\text{Im}\Sigma(\bm k,\epsilon_{\bm k})|}{k|\partial_{\bm k} \epsilon_{\bm k}|} = (N+1)\pi s^2 \frac{k^2+k_{\star}^2}{k_{\star}^2}.
\end{equation}
The quasiparticles are well resolved in momentum in the domain $k\lesssim k_\star/\sqrt{(N+1)\pi s^2}$. For a sufficiently small $s$, this covers the most interesting dispersive part of the spectrum in Eq.~\eqref{eq:twomodes}.
We note that in contrast, the states of the upper and lower Hubbard bands are {\sl not resolved} in momentum in the spin-disordered $U/t\gg 1$ Hubbard model~\cite{hubbard1964III}.

The relaxation rates for the quasiparticle spectra in Eq.~\eqref{eq:fourmodes} for $M\neq 0$ are
\begin{align}\label{eq:rate_M}
    \frac{1}{\tau_{\bm k\pm}^{i}} &\equiv 2 Z_{\bm k\pm}^{i}|\text{Im}\Sigma(\bm k,\epsilon_{\bm k\pm}^{i})|\notag\\
    &\approx (N+1)\pi s^2\frac{\frac{U^2}{4}\frac{k^2}{k^2+k_{\star}^2}}{\sqrt{\frac{M^2}{4}\big(\frac{k_{\star}^2}{k^2+k_{\star}^2}\big)^2+\frac{U^2}{4}\frac{k^2}{k^2+k_{\star}^2}}}\notag\\
    &\quad \times\left[1 + \frac{M-|\epsilon_{\bm k\pm}^{i}|}{2|\epsilon_{\bm k\pm}^{i}|}\theta(M-|\epsilon_{\bm k\pm}^{i}|)\right],\, i=1,2.
\end{align}
The relaxation rates for all branches approach the form of Eq.~(\ref{eq:rateMzero}) for $k\gg(M/U)k_\star$. The modes $\epsilon_{\bm k\pm}^{2}$ remain well-defined down to $k\to 0$, where their decay rates scale quadratically with $k$,
\begin{equation}\label{eq:ralaxation_electron}
\frac{1}{\tau_{\bm k\pm}^{2}}\approx (N+1)\pi s^2\frac{U^2}{2M}\frac{k^2}{k_\star^2},\ \ \ k\ll \frac{M}{U}k_{\star},
\end{equation}
while $\epsilon_{\bm k\pm}^{2}\to \pm M$.   
Interestingly, the trion relaxation rate remains finite at the Fermi level, 
\begin{equation}\label{eq:relaxation_trion}
    \frac{1}{\tau_{0\pm}^1}\approx(N+1)\pi s^2 M.
\end{equation}
As $k$ approaches a small domain around the $\Gamma$-point, $k\sim \sqrt{4\pi N}sMk_{\star}/U$, the relaxation rate $1/\tau_{\bm k\pm}^1\sim |\epsilon_{\bm k\pm}^1|$.
\begin{figure}
    \centering
    \includegraphics[width=1\linewidth]{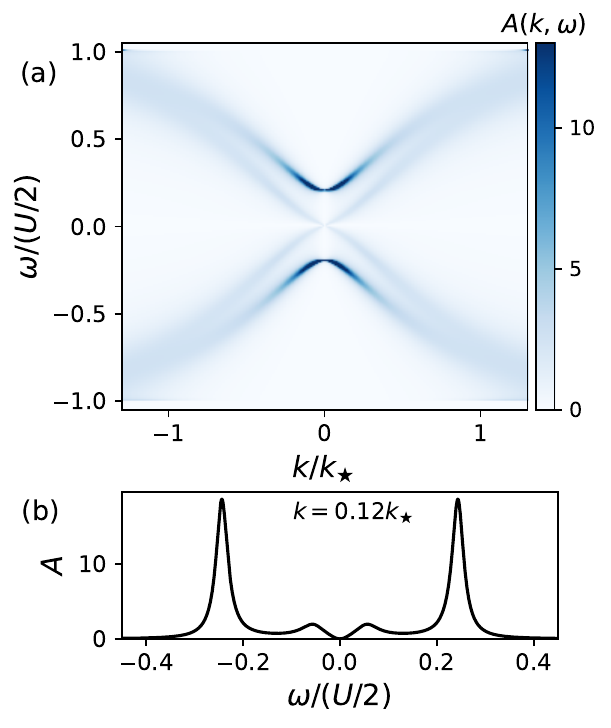}
    \caption{(a) The Hubbard-III spectral function $A(\bm k,\omega)$ of the projected THF model as a function of momentum $k$ and energy $\omega$. For this plot, we used the self-energy, Eqs.~\eqref{eq:self-energy_re_HubbardIIIM} and~\eqref{eq:self-energy_im_HubbardIIIM}, expanded to first-order in $s^2$ with $s=0.1$ and $M=0.1U$. (b) An energy distribution curve along $k=0.12k_{\star}$. The two prominent peaks are single-particle excitations at energies$\sim \pm M$, and the two weaker peaks closer to the Fermi level correspond to trions.
    }
    \label{fig:spectra}
\end{figure}

Having found the self-energy [see Eqs.~\eqref{eq:self-energy_HubbardIIIM}], we may evaluate the electron spectral function,
\begin{equation}
    A(\bm k,\omega) = -\frac{1}{\pi}\sum_{\alpha=1,2}\text{Im}\ {\langle\langle d_{\bm k\eta \alpha};d_{\bm k\eta \alpha}^{\dagger}\rangle\rangle_\omega}.
\end{equation}
Its behavior around the $\Gamma$-point is presented in Fig.~\ref{fig:spectra}(a). We see that the two branches of high-energy excitations have large residue and narrow linewidth at small $k$. Figure~\ref{fig:spectra}(b) is a vertical line cut at $k=0.12k_{\star}$. While two weak trion peaks remain observable at this momentum, their energies become comparable to the relaxation rate, Eq.~\eqref{eq:relaxation_trion}, and the peak positions start to shift away from the values $\epsilon_{\bm k}$ determined by $\epsilon_{\bm k} -\text{Re}\Sigma_{\text{III}}(\bm k,\epsilon_{\bm k})=0$.

\section{Real-space Hamiltonian and Hubbard-I approximation}

We now turn to the details of the derivation, starting with the Hubbard-I approximation in this section. To treat the strong local correlations resulting from the on-site repulsion $H_U$, we find it convenient to introduce a localized basis for the TBG flat bands by projecting the $f$ orbitals onto the flat bands, 
\begin{equation}\label{eq:tildef_R}
    \tilde{f}_{i \sigma} = \frac{1}{\sqrt{z}}Pf_{i}P=\frac{1}{\sqrt{N_s z}}\sum_{\bm k} e^{i\bm k\cdot\bm R_i}\frac{k}{\sqrt{k^2+k_{\star}^2}}d_{\bm k \sigma},
\end{equation}
with normalization constant
\begin{equation}\label{eq:z}
    z = \int_{\text{mBZ}}\frac{d^2k}{\Omega_m} \frac{k^2}{k^2+k_{\star}^2} \approx 1 - s^2 \ln \frac{1}{s^2}.
\end{equation}
The operators $\tilde{f}_{i\sigma}$ constitute a nonorthogonal basis distinct from the Wannier basis of the flat bands, and obey the anticommutation relations
\begin{align}
    &\{\tilde{f}_{i \sigma}, \tilde{f}_{i' \sigma'} \} = 0,\\
    &\{\tilde{f}_{i \sigma}, \tilde{f}_{i' \sigma'}^{\dagger} \} = \delta_{\sigma,\sigma'} \lambda_{ii'},\label{eq:commutator}
\end{align}
where $\lambda_{ij}$ [see Eq.~\eqref{eq:lambda}] satisfies $\lambda_{ij}= \lambda_{ji}^* = \lambda_{ji}$. 
%
%

In terms of the operators $\tilde{f}_{i\sigma}$,
we obtain a compact real-space expression for the projected interaction in Eq.~\eqref{eq:PHP},
\begin{equation}\label{eq:H_U_project}
    \tilde{H}_U = \frac{U}{2}\sum_{i}\left(\sum_{\sigma=1}^{N}\tilde{f}_{i \sigma}^{\dagger}\tilde{f}_{i \sigma} -\frac{N}{2}\right)^2.
\end{equation}
Strictly speaking, the interaction strength is reduced to $Uz^2$, but we have neglected the rescaling factor $z^2\approx 1$.   

Equation~\eqref{eq:H_U_project} corresponds to a Hubbard interaction on a non-orthogonal basis. Unlike the conventional Hubbard interaction, $\tilde{H}_U$ can transfer electrons between different sites,
\begin{equation}\label{eq:commutator_fH}
    [\tilde{f}_{i \sigma}, \tilde{H}_U] = U\sum_{i'}\lambda_{ii'}\delta n_{i' \bar{\sigma}}\tilde{f}_{i' \sigma},
\end{equation}
with
\begin{equation}
    \delta n_{i \bar{\sigma}}\equiv \sum_{\sigma'\neq \sigma}\left(\tilde{f}_{i \sigma'}^{\dagger}\tilde{f}_{i\sigma'}-\frac{1}{2}\right).
\end{equation}
Thus, $\tilde H_U$ can give rise to a finite dispersion of electrons, even if the band is flat in the absence of interactions ($M=0$ limit).

In this work, we study the single-particle Green's function of the projected THF model at charge neutrality using the equation of motion method \cite{zubarev1960}. We denote the retarded Green's function of two operators $A$ and $B$, which are odd under fermion parity, as
\begin{align}
  \langle\langle A; B\rangle\rangle \equiv-i \theta(t) \langle \left\{A(t), B(0)\right\} \rangle.
\end{align}
Here, $\theta(t)$ is the Heaviside function, and the angular bracket stands for ensemble averaging.
The Green's function satisfies the equation of motion
\begin{equation}\label{eq:eom}
    i\partial_t\langle\langle A; B\rangle\rangle = \delta(t)\langle \{A(0), B(0)\}\rangle +\langle\langle [A,\tilde{H}]; B\rangle\rangle.
\end{equation}

We focus first on the chiral-flat limit, $M=0$. Using the relation Eq.~\eqref{eq:commutator_fH} and the shorthand notation $\psi_{i\sigma} \equiv \delta n_{i\bar{\sigma}}\tilde{f}_{i\sigma}=\tilde{f}_{i\sigma}\delta n_{i\bar{\sigma}}$ for the `trion' operator \cite{ledwith2025exotic}, we derive the equations of motion
\begin{align}
    i\partial_t\langle\langle \tilde{f}_{j \sigma}; \tilde{f}_{i \sigma}^{\dagger}\rangle\rangle &= \lambda_{ji}\delta(t) + U\sum_{l} \lambda_{jl} \langle\langle \psi_{l \sigma}; \tilde{f}_{i \sigma}^{\dagger}\rangle\rangle, \label{eq:eom_f}\\
    i\partial_t\langle\langle \psi_{j \sigma}; \tilde{f}_{i \sigma}^{\dagger}\rangle\rangle &= \lambda_{ji}\langle \delta n_{j\bar{\sigma}}\rangle\delta(t) + U\langle\langle \delta n_{j\bar{\sigma}}^2\tilde{f}_{j \sigma}; \tilde{f}_{i  \sigma}^{\dagger}\rangle\rangle\notag\\
    &\quad+U\sum_{l\neq j}\lambda_{jl}\langle\langle \delta n_{j\bar{\sigma}}\delta n_{l\bar{\sigma}}\tilde{f}_{l \sigma}; \tilde{f}_{i  \sigma}^{\dagger}\rangle\rangle\notag\\
    &\quad+\langle\langle [\delta n_{j\bar{\sigma}},\tilde{H}]\tilde{f}_{l \sigma}; \tilde{f}_{i  \sigma}^{\dagger}\rangle\rangle.\label{eq:eom_psi}
\end{align}
%
We restrict our discussion to the paramagnetic Mott state in which the charge is frozen, $\sum_{\sigma}f_{j\sigma}^\dagger f_{j\sigma}=N/2$, but flavor is thermally disordered, i.e., $\langle f_{j\sigma}^\dagger f_{j\sigma}\rangle$ is independent of $\sigma$. This allows us to simplify the first line of Eq.~\eqref{eq:eom_psi} using 
%
\begin{subequations}
\begin{equation}\label{eq:nopolarization}
    \langle \delta n_{j\bar{\sigma}}\rangle=0,\quad \langle \delta n_{j\bar{\sigma}}^2\rangle\approx\frac{1}{4},
\end{equation}
and
\begin{equation}\label{eq:factorization}
    \langle\langle\delta n_{j\bar{\sigma}}^2f_{j\sigma};f_{i\sigma}^{\dagger}\rangle\rangle\approx \langle \delta n_{j\bar{\sigma}}^2\rangle\langle\langle f_{j\sigma};f_{i\sigma}^{\dagger}\rangle\rangle.
\end{equation}
\end{subequations}
%
Relations~\eqref{eq:nopolarization} and~\eqref{eq:factorization} become exact in the atomic limit ($s\to 0$), or at any $s$ in case of $N=2$.
The Green's functions in the second and third lines of Eq.~\eqref{eq:eom_psi} both involve two different sites at the same time $t$, and hence are suppressed by a factor of $s^2$. 

To the zeroth order in $s$, we retain only the first line of Eq.~\eqref{eq:eom_psi}, and simplify it by using Eqs.~(\ref{eq:nopolarization}), (\ref{eq:factorization}). This corresponds to the Hubbard-I approximation \cite{hubbard1963I}, and leads to a closed system of equations for the Green's function. 
Fourier transforming,
\begin{equation}
\langle\langle A; B\rangle\rangle_{\omega} \equiv \int_{-\infty}^{+\infty} d t\ \langle\langle A; B\rangle\rangle e^{i \omega t},
\end{equation}
and solving Eqs.~(\ref{eq:eom_f}) and (\ref{eq:eom_psi}), we find the Green's function
\begin{equation}
    \langle\langle d_{\bm k \sigma}; d_{\bm k \sigma}^{\dagger}\rangle\rangle_{\omega} = \frac{1}{\omega - \Sigma(\bm k,\omega)},\label{eq:g_dd_k}
\end{equation}
with the self-energy in Hubbard-I approximation, 
\begin{equation}
\Sigma_{\rm I}(\bm k, \omega) = \frac{U^2}{4\omega}\frac{{k^2}}{({k^2+k_{\star}^2})}.    
\end{equation}
The poles of the Green's function in Eq.\ \eqref{eq:g_dd_k} determine the quasiparticle spectra in Eq.~\eqref{eq:twomodes}. 

In our derivation, the dispersion of the Hubbard bands arises from the nonlocal orbital overlap, $\lambda_{ij}(i\neq j)$. Although each individual $\lambda_{ij}\sim s^2$, their cumulative contribution remains of order unity in Eq.~\eqref{eq:eom_f}, as each orbital overlaps with orbitals on $\sim 1/s^2$ sites.

\section{Hubbard-III approximation}
In this section, we evaluate the right-hand side of Eq.~\eqref{eq:eom_psi} to the next order in $s^2$ by truncating the equations-of-motion hierarchy following Ref.~\cite{hubbard1964III}, which we refer to as the Hubbard-III approximation. Unlike in the Hubbard model, this approximation becomes controlled in our case by the small parameter $s^2$.
For simplicity, we focus on the chiral-flat limit $M=0$ in this section. 

In line with the original Hubbard-III approximation~\cite{hubbard1964III}, we separate the emerging correction into scattering and resonance broadening contributions.  These arise from the two last terms on the right-hand side of Eq.~(\ref{eq:eom_psi}). To order $s^2$, the contributions are additive in the self-energy and can therefore be  considered separately.
\subsection{The scattering correction}\label{sec:scattering_correction}
The equation of motion for the Green's function appearing in the second line of Eq.~\eqref{eq:eom_psi} reads
\begin{equation}\label{eq:eom_nnf}
\begin{split}
         &i\partial_t \langle\langle \delta n_{j\bar{\sigma}}\delta n_{l\bar{\sigma}}\tilde{f}_{l\sigma}; \tilde{f}_{i \sigma}^{\dagger}\rangle\rangle \\
        =&\ \lambda_{li}\langle \delta n_{j\bar{\sigma}}\delta n_{l\bar{\sigma}}\rangle \delta(t)+ U\langle\langle \delta n_{j\bar{\sigma}}\delta n_{l\bar{\sigma}}^2\tilde{f}_{l\sigma}; \tilde{f}_{i \sigma}^{\dagger}\rangle\rangle\\
        &+ U\sum_{r\neq l}\lambda_{lr}\langle\langle \delta n_{j\bar{\sigma}}\delta n_{l\bar{\sigma}}\delta n_{r\bar{\sigma}}\tilde{f}_{r\sigma}; \tilde{f}_{i \sigma}^{\dagger}\rangle\rangle\\
        &- \langle\langle \big[\tilde{H}, \delta n_{j\bar{\sigma}}\delta n_{l\bar{\sigma}}\big]\tilde{f}_{l\sigma}; \tilde{f}_{i \sigma}^{\dagger}\rangle\rangle.
\end{split}
\end{equation}
Here, $j\neq l$ and therefore $\langle \delta n_{j\bar{\sigma}}\delta n_{l\bar{\sigma}}\rangle\sim O(s^4)$ \footnote{Because two fermionic operators at different sites have $O(s^2)$ overlap, we expect that the correlation between two density operators at different sites, each of which is composed of a pair of fermionic operators, should be $O(s^4)$} is negligible. 
All remaining terms on the right-hand side of Eq.~\eqref{eq:eom_nnf} are proportional to $U$ and contain 
a larger number of the fermionic operators than the left-hand side. To truncate the chain of equations of motion, we first drop the third and fourth terms as their Green's functions generically involve three different sites at the same time $t$ \footnote{The Green's functions at $r=j$ is the only exception, which contains two different sites at the same time $t$. However, this single case can be neglected due to its small weight in the summation $(\lambda_{lj}\sim O(s^2))$.} and are of a higher order in $s^2$ than the second term. Next, we may also factorize the second term in the same way as in Eq.~\eqref{eq:factorization}.
Thus, Eq.~\eqref{eq:eom_nnf} simplifies to
\begin{equation}\label{eq:eom_nnf_approx}
    i\partial_t \langle\langle \delta n_{j\bar{\sigma}}\delta n_{l\bar{\sigma}}\tilde{f}_{l\sigma}; \tilde{f}_{i \sigma}^{\dagger}\rangle\rangle = U\langle\delta n_{l\bar{\sigma}}^2\rangle\langle\langle \delta n_{j\bar{\sigma}}\tilde{f}_{l\sigma}; \tilde{f}_{i \sigma}^{\dagger}\rangle\rangle.
\end{equation}
The Green's function on the right-hand side of the equation now has fewer fermionic operators. However, it remains distinct from those  appearing in Eqs.~\eqref{eq:eom_f} and~\eqref{eq:eom_psi}. To close the equations, we write down the equation of motion for the Green's function appearing in the right-hand side of Eq.~(\ref{eq:eom_nnf_approx}),
\begin{equation}\label{eq:eom_nf}
    i\partial_t \langle\langle \delta n_{j\bar{\sigma}}\tilde{f}_{l\sigma}; \tilde{f}_{i \sigma}^{\dagger}\rangle\rangle = U\sum_{r}\lambda_{lr} \langle\langle \delta n_{j\bar{\sigma}}\delta n_{r\bar{\sigma}}\tilde{f}_{r\sigma}; \tilde{f}_{i \sigma}^{\dagger}\rangle\rangle.
\end{equation}
At $r\neq j$, the Green's functions on the right-hand side coincide with those on the left-hand side of Eq.~\eqref{eq:eom_nnf_approx}. At $r=j$, we can factorize the Green's functions on the right-hand side of Eq.~\eqref{eq:eom_nf} according to Eq.~\eqref{eq:factorization}.

Combining Eqs.~\eqref{eq:eom_nnf_approx} and~\eqref{eq:eom_nf} and Fourier transforming into frequency space, we arrive at
\begin{equation}
\begin{split}
    \omega \langle\langle \delta n_{j\bar{\sigma}}\tilde{f}_{l\sigma}; \tilde{f}_{i \sigma}^{\dagger}\rangle\rangle_{\omega} &- \sum_{r\neq j}t_{lr}^{\scriptscriptstyle U}(\omega) \langle\langle \delta n_{j\bar{\sigma}}\tilde{f}_{r\sigma}; \tilde{f}_{i \sigma}^{\dagger}\rangle\rangle_{\omega}\\
    &\quad = \frac{\omega}{U}t_{lj}^{\scriptscriptstyle U}(\omega)  \langle\langle \tilde{f}_{j \sigma}; \tilde{f}_{i \sigma}^{\dagger}\rangle\rangle_{\omega} \label{eq:eom_nf_freq}
\end{split}
\end{equation}
with an interaction-induced hopping amplitude
\begin{equation}
    t_{lj}^{\scriptscriptstyle U}(\omega) = \frac{U^2\langle\delta n_{j\bar{\sigma}}^2\rangle}{\omega}\lambda_{lj}.
\end{equation}
A method to solve this type of linear equation was presented in the original paper introducing the Hubbard-III approximation \cite{hubbard1964III}. For completeness, we provide the derivation in Appendix~\ref{sec:W}. The solution of Eq.~(\ref{eq:eom_nf_freq}) is
%
\begin{align}
    \langle\langle \delta n_{j\bar{\sigma}}\tilde{f}_{l\sigma}; \tilde{f}_{i \sigma}^{\dagger}\rangle\rangle_{\omega} = \frac{\omega}{U}\sum_{r\neq j}W_{lr, j}t_{rj}^{\scriptscriptstyle U} \langle\langle \tilde{f}_{j \sigma}; \tilde{f}_{i \sigma}^{\dagger}\rangle\rangle_{\omega}.\label{eq:nf_freq}
\end{align}
Here, we introduce 
\begin{align}
    W_{lr, j}(\omega) =  g_{lr}(\omega) - \frac{g_{lj}(\omega)g_{jr}(\omega)}{g_{jj}(\omega)}, \label{eq:W}
\end{align}
%
with the Green's function $g_{jl}(\omega)=[\omega - \hat{t}^{\scriptscriptstyle U}(\omega)]_{jl}^{-1}$. Note that the Fourier transform of $g_{jl}$ in momentum space corresponds to the Hubbard-I Green's function
\begin{equation}
    g(\bm k,\omega)= \sum_{i} g_{ij} e^{i\bm k\cdot(\bm R_j-\bm R_i)}= \frac{1}{\omega - \frac{U^2}{4\omega}\frac{k^2}{k^2+k_{\star}^2}}.\label{eq:g_ff_k}
\end{equation}
%

Plugging Eq.~\eqref{eq:nf_freq} into the Fourier transform of Eq.~\eqref{eq:eom_nnf_approx} and using the relation
\begin{equation*}\label{eq:eom_psi_HubbardI}
\omega\langle\langle\psi_{j\bar{\sigma}};f_{i\sigma}^{\dagger}\rangle\rangle_{\omega} \approx U\langle\delta n_{j\bar{\sigma}}^2\rangle\langle\langle f_{j\bar{\sigma}};f_{i\sigma}^{\dagger}\rangle\rangle_{\omega},
\end{equation*}
valid within the accuracy of Hubbard-III approximation we obtain
\begin{equation}
    \langle\langle \delta n_{j\bar{\sigma}}\delta n_{l\bar{\sigma}}\tilde{f}_{l\sigma}; \tilde{f}_{i \sigma}^{\dagger}\rangle\rangle_{\omega} = \frac{\omega}{U}\sum_{r\neq j}W_{lr, j}t_{rj}^{\scriptscriptstyle U}\langle\langle \psi_{j \sigma}; \tilde{f}_{i \sigma}^{\dagger}\rangle\rangle_{\omega}.\label{eq:nnf_freq}
\end{equation}
This expression allows us to compute the scattering correction in Eq.~\eqref{eq:eom_psi},
\begin{equation}\label{eq:scattering_correction}
\begin{split}
    &U\sum_{l\neq j}\lambda_{jl}\langle\langle \delta n_{j\bar{\sigma}}\delta n_{l\bar{\sigma}}\tilde{f}_{l\sigma}; \tilde{f}_{i \sigma}^{\dagger}\rangle\rangle_{\omega}= c(\omega) \omega \langle\langle\psi_{j\sigma}; \tilde{f}_{i \sigma}^{\dagger}\rangle\rangle_{\omega}.
\end{split}
\end{equation}
The coefficient $c(\omega)$ reads (see Appendix~\ref{sec:W})
\begin{subequations}
\begin{align}
    c(\omega) &= \frac{\sum_{\bm k} g(\bm k,\omega)\bar{\lambda}(\bm k)}{\sum_{\bm k}g(\bm k,\omega)} \label{eq:c1_def}\\
    &\approx -s^2\ln{\Big(1-\frac{U^2}{ 4\omega^2}\Big)},\label{eq:c1_approx}
\end{align}
\end{subequations}
where $\bar{\lambda}(\bm k) = \sum_{i\neq j} \lambda_{ij} e^{-i\bm k\cdot (\bm R_i-\bm R_j)}$ denotes the Fourier transform of the intersite orbital overlap $\lambda_{ij} (i\neq j)$. The last equation is derived by expanding Eq.~\eqref{eq:c1_def} to leading order in $s^2$. The complex logarithmic function has a branch cut along the negative real axis. Its imaginary part ultimately contributes to a finite relaxation rate.

\subsection{The resonance broadening correction}
\label{sec:resonance}
The resonance broadening correction arises from the last term on the right-hand side of Eq.~(\ref{eq:eom_psi}),
\begin{equation}\label{eq:resonance_broadening}
\begin{split}
    &\langle\langle [\delta n_{j\bar{\sigma}},\tilde{H}]\tilde{f}_{l \sigma}; \tilde{f}_{i  \sigma}^{\dagger}\rangle\rangle \\
    =&\ U\sum_{l\neq j}\sum_{\sigma'\neq \sigma}\lambda_{lj}\langle\langle \delta n_{l \overline{\sigma'}}\big(\tilde{f}_{j \sigma'}^{\dagger}\tilde{f}_{l \sigma'}-\tilde{f}_{l \sigma'}^{\dagger}\tilde{f}_{j \sigma'}\big)\tilde{f}_{j \sigma}; \tilde{f}_{i \sigma}^{\dagger}\rangle\rangle.
\end{split}
\end{equation}
Similar to the evaluation of the scattering correction, we need to write the equations of motion for the correlators appearing here, and then truncate the chain of equations by factorizing the higher-order correlation functions appearing in those equations.
In this section, we perform this program for $N=2$, leaving the more cumbersome analysis of a general even $N$ to Appendix~\ref{sec:evenN}.

In evaluating the Green's functions on the right-hand side of Eq.~\eqref{eq:resonance_broadening}, we expand the parenthesis and treat the two resulting terms separately. The first term obeys the equation of motion
\begin{align}\label{eq:eom_fpsif}
        & i\partial_t\langle\langle \tilde{f}_{j \sigma'}^{\dagger}\delta n_{l \overline{\sigma'}}\tilde{f}_{l \sigma'}\tilde{f}_{j \sigma}; \tilde{f}_{i \sigma}^{\dagger}\rangle\rangle  \notag\\
        \approx &\ \lambda_{ji}\langle \tilde{f}_{j \sigma'}^{\dagger}\delta n_{l \overline{\sigma'}}\tilde{f}_{l \sigma'}\rangle \delta(t)
        + U\langle\langle \tilde{f}_{j \sigma'}^{\dagger}\delta n_{l \overline{\sigma'}}^2\tilde{f}_{l \sigma'}\tilde{f}_{j \sigma}; \tilde{f}_{i \sigma}^{\dagger}\rangle\rangle \notag\\
        &- U\sum_{r\neq j} \lambda_{rj}\langle\langle \delta n_{r \overline{\sigma'}}\tilde{f}_{r \sigma'}^{\dagger}\delta n_{l \overline{\sigma'}}\tilde{f}_{l \sigma'}\tilde{f}_{j \sigma}; \tilde{f}_{i \sigma}^{\dagger}\rangle\rangle \notag\\
        &+ U\sum_{r\neq j} \lambda_{jr}\langle\langle \tilde{f}_{j \sigma'}^{\dagger}\delta n_{l \overline{\sigma'}}\tilde{f}_{l \sigma'}\delta n_{r \bar{\sigma}}\tilde{f}_{r \sigma}; \tilde{f}_{i \sigma}^{\dagger}\rangle\rangle.
\end{align}
%
%
We remind the reader that $\bar\sigma\neq\sigma$. On the right-hand side, we made approximations by (i) dropping a spin-exchange correlation function $\lambda_{li}\langle \tilde{f}_{j \sigma'}^{\dagger}\tilde{f}_{l \sigma}^{\dagger}\tilde{f}_{l \sigma'}\tilde{f}_{j \sigma}\rangle \delta(t)$ in the first term; (ii) keeping only the on-site scattering in $[\tilde{f}_{l\sigma},\tilde{H}_U]$ in the second term; and (iii) neglecting $[\delta n_{l\overline{\sigma'}}, \tilde{H}_U]$. These omitted terms would result in corrections which are of higher order in $s^2$. In the last two lines of Eq.~(\ref{eq:eom_fpsif}), the Green's functions at $r=j$ cancel. 

The equation of motion for the second part of the Green's functions in Eq.~\eqref{eq:resonance_broadening} can be derived in a similar manner,
\begin{align}
        & i\partial_t\langle\langle \delta n_{l \overline{\sigma'}}\tilde{f}_{l \sigma'}^{\dagger}\tilde{f}_{j \sigma'}\tilde{f}_{j \sigma}; \tilde{f}_{i \sigma}^{\dagger}\rangle\rangle  \notag\\
        \approx &\ \lambda_{ji}\langle \delta n_{l \overline{\sigma'}}\tilde{f}_{l \sigma'}^{\dagger}\tilde{f}_{j \sigma'}\rangle\delta(t) - U\langle\langle \delta n_{l \overline{\sigma'}}^2\tilde{f}_{l \sigma'}^{\dagger}\tilde{f}_{j \sigma'}\tilde{f}_{j \sigma}; \tilde{f}_{i \sigma}^{\dagger}\rangle\rangle \notag\\
        &+ U\sum_{r\neq j} \lambda_{jr}\langle\langle \delta n_{l \overline{\sigma'}}\tilde{f}_{l \sigma'}^{\dagger}\delta n_{r \overline{\sigma'}}\tilde{f}_{r \sigma'}\tilde{f}_{j \sigma}; \tilde{f}_{i \sigma}^{\dagger}\rangle\rangle \notag\\
        &+ U\sum_{r\neq j} \lambda_{rj}\langle\langle \delta n_{l \overline{\sigma'}}\tilde{f}_{l \sigma'}^{\dagger}\tilde{f}_{j \sigma'}\delta n_{r \bar{\sigma}}\tilde{f}_{r \sigma}; \tilde{f}_{i \sigma}^{\dagger}\rangle\rangle.
        \label{eq:eom_psiff}
\end{align}
In the third and fourth terms, the Green's functions at $r=j$ cancel. This can be shown by noticing that $\delta n_{l\bar{\sigma}}=f_{l\sigma'}^{\dagger}f_{l\sigma'}-1/2$ and $\delta n_{l\overline{\sigma'}}=f_{l\sigma}^{\dagger}f_{l\sigma}-1/2$ for $N=2$. For general $N$, the cancellation is not exact and the corrections are analyzed in Appendix~\ref{sec:evenN}.

Next, we truncate the equations of motion by factorizing Green's functions as follows. In the third term on the right-hand side of Eq.~\eqref{eq:eom_fpsif},
\begin{equation}
\begin{split}\label{eq:factorization2}
        &\langle\langle \delta n_{r \overline{\sigma'}}\tilde{f}_{r \sigma'}^{\dagger}\delta n_{l \overline{\sigma'}}\tilde{f}_{l \sigma'}\tilde{f}_{j \sigma}; \tilde{f}_{i \sigma}^{\dagger}\rangle\rangle\\
        \approx &\ \delta_{l,r}\langle\delta n_{l \overline{\sigma'}}^2\rangle\langle\tilde{f}_{l \sigma'}^{\dagger}\tilde{f}_{l \sigma'}\rangle\langle\langle \tilde{f}_{j \sigma}; \tilde{f}_{i \sigma}^{\dagger}\rangle\rangle, \\
\end{split}
\end{equation}
because $j\neq r,l$. In the fourth term, the Green’s functions generally involve three different sites \footnote{The single exception at $r=l$ can be neglected because the factor $\lambda_{jl}$ suppress its contribution by $O(s^2)$ compared with the remaining terms. } and can be decoupled into
\begin{equation}
\begin{split}\label{eq:factorization3}
    &\langle\langle \tilde{f}_{j \sigma'}^{\dagger}\delta n_{l \overline{\sigma'}}\tilde{f}_{l \sigma'}\delta n_{r \bar{\sigma}}\tilde{f}_{r \sigma}; \tilde{f}_{i \sigma}^{\dagger}\rangle\rangle \\
    \approx\ & \langle \tilde{f}_{j \sigma'}^{\dagger}\delta n_{l \overline{\sigma'}}\tilde{f}_{l \sigma'}\rangle\langle\langle \delta n_{r \bar{\sigma}}\tilde{f}_{r \sigma}; \tilde{f}_{i \sigma}^{\dagger}\rangle\rangle.\\
\end{split}
\end{equation}
The Green's functions in the last two lines of Eq.~\eqref{eq:eom_psiff} can be factorized similarly. The factorization result can be brought to a form identical to that of 
the right-hand sides of Eqs.~\eqref{eq:factorization2} and~\eqref{eq:factorization3} upon using the relation $\langle \tilde{f}_{j \sigma'}^{\dagger}\delta n_{l \overline{\sigma'}}\tilde{f}_{l \sigma'}\rangle = \langle \delta n_{l \overline{\sigma'}}\tilde{f}_{l \sigma'}^{\dagger}\tilde{f}_{j \sigma'}\rangle$, see Appendix~\ref{sec:gf}. 

%

As a consequence, subtracting Eq.~\eqref{eq:eom_psiff} from Eq.~\eqref{eq:eom_fpsif} cancels the first and fourth terms on the right-hand side exactly. The result reads
\begin{align}
        & i\partial_t\langle\langle \delta n_{l \overline{\sigma'}}\big(\tilde{f}_{j \sigma'}^{\dagger}\tilde{f}_{l \sigma'}-\tilde{f}_{l \sigma'}^{\dagger}\tilde{f}_{j \sigma'}\big)\tilde{f}_{j \sigma}; \tilde{f}_{i \sigma}^{\dagger}\rangle\rangle  \notag\\
        \approx & U\langle\delta n_{l\overline{\sigma'}}^2\rangle \langle\langle \big(\tilde{f}_{j \sigma'}^{\dagger}\tilde{f}_{l \sigma'}+\tilde{f}_{l \sigma'}^{\dagger}\tilde{f}_{j \sigma'}\big) \tilde{f}_{j \sigma}; \tilde{f}_{i \sigma}^{\dagger}\rangle\rangle^{c}. \label{eq:eom_fpsif-psiff}
\end{align}
Here, we introduced the connected Green's function
\begin{equation}\label{eq:gf_connected}
    \langle\langle \hat{O}_j\tilde{f}_{j \sigma}; \tilde{f}_{i \sigma}^{\dagger}\rangle\rangle^{c}\equiv \langle\langle \hat{O}_j\tilde{f}_{j \sigma}; \tilde{f}_{i \sigma}^{\dagger}\rangle\rangle - \langle\hat{O}_j\rangle\langle\langle \tilde{f}_{j \sigma}; \tilde{f}_{i \sigma}^{\dagger}\rangle\rangle,
\end{equation}
where $\hat{O}_j$ is a product of fermion operators on site $j$ and other sites.
Specifically, we used $\langle\tilde{f}_{j \sigma'}^{\dagger}\tilde{f}_{l \sigma'} \rangle = \langle\tilde{f}_{l \sigma'}^{\dagger}\tilde{f}_{j \sigma'} \rangle = \lambda_{jl}\langle \tilde{f}_{l\sigma'}^{\dagger}\tilde{f}_{l\sigma'}\rangle$ at charge neutrality $\nu=0$, see Appendix~\ref{sec:gf}.

To compute the connected Green's functions, we derive their equations of motion from Eq.~\eqref{eq:eom},
%
%
%
\begin{equation}\label{eq:eom_fff1}
\begin{split}
    &i\partial_t\langle\langle \tilde{f}_{j \sigma'}^{\dagger}\tilde{f}_{l \sigma'}\tilde{f}_{j \sigma}; \tilde{f}_{i \sigma}^{\dagger}\rangle\rangle^{c} \\
    \approx&\ \frac{U}{4}\lambda_{lj}\langle\langle \tilde{f}_{j \sigma}; \tilde{f}_{i \sigma}^{\dagger}\rangle\rangle\\ 
    &+ U\sum_{r\neq j}\lambda_{jr}\langle\langle (\tilde{f}_{j \sigma'}^{\dagger}\tilde{f}_{l \sigma'}-\langle\tilde{f}_{j \sigma'}^{\dagger}\tilde{f}_{l \sigma'}\rangle)\psi_{r \sigma}; \tilde{f}_{i \sigma}^{\dagger}\rangle\rangle\\
    &-U\sum_{r\neq j}\lambda_{jr}\langle\langle \delta n_{r\overline{\sigma'}}\tilde{f}_{r \sigma'}^{\dagger}\tilde{f}_{l \sigma'}\tilde{f}_{j \sigma}; \tilde{f}_{i \sigma}^{\dagger}\rangle\rangle\\
    & +U\sum_{r\neq j}\lambda_{lr}\langle\langle \tilde{f}_{j \sigma'}^{\dagger}\delta n_{r\overline{\sigma'}}\tilde{f}_{r \sigma'}\tilde{f}_{j \sigma}; \tilde{f}_{i \sigma}^{\dagger}\rangle\rangle.
\end{split}
\end{equation}
The missing $r=j$ terms in the sums appearing in the last three lines here are combined (for $N=2$) with the first two terms on the right-hand side of Eq.~(\ref{eq:eom_fff1}).
Similarly, 
\begin{equation}\label{eq:eom_fff2}
\begin{split}
    &i\partial_t\langle\langle \tilde{f}_{l \sigma'}^{\dagger}\tilde{f}_{j \sigma'}\tilde{f}_{j \sigma}; \tilde{f}_{i \sigma}^{\dagger}\rangle\rangle^{c} \\
    \approx&\ \frac{U}{4}\lambda_{lj}\langle\langle \tilde{f}_{j \sigma}; \tilde{f}_{i \sigma}^{\dagger}\rangle\rangle\\ 
    &+ U\sum_{r\neq j}\lambda_{jr}\langle\langle (\tilde{f}_{l \sigma'}^{\dagger}\tilde{f}_{j \sigma'}-\langle\tilde{f}_{l \sigma'}^{\dagger}\tilde{f}_{j \sigma'}\rangle)\psi_{r \sigma}; \tilde{f}_{i \sigma}^{\dagger}\rangle\rangle\\
    & +U\sum_{r\neq j}\lambda_{jr}\langle\langle \tilde{f}_{l \sigma'}^{\dagger}\delta n_{r\overline{\sigma'}}\tilde{f}_{r \sigma'}\tilde{f}_{j \sigma}; \tilde{f}_{i \sigma}^{\dagger}\rangle\rangle\\
    &-U\sum_{r\neq j}\lambda_{rl}\langle\langle \delta n_{r\overline{\sigma'}}\tilde{f}_{r \sigma'}^{\dagger}\tilde{f}_{j \sigma'}\tilde{f}_{j \sigma}; \tilde{f}_{i \sigma}^{\dagger}\rangle\rangle.
\end{split}
\end{equation}
%
%
Next, we analyze the higher-order correlation functions on the right-hand sides of Eqs.~\eqref{eq:eom_fff1} and~\eqref{eq:eom_fff2}, i.e., the three last lines in each of these equations. Correlation functions in the first two of the three lines contain equal-time ($t$) creation and annihilation operators at three different sites; therefore, these functions 
can be factorized to the desired accuracy. Upon factorization, the first of the three lines turns to zero. The second one yields
%
\begin{equation*}
    \langle\langle \delta n_{r\overline{\sigma'}}\tilde{f}_{r \sigma'}^{\dagger}\tilde{f}_{l \sigma'}\tilde{f}_{j \sigma}; \tilde{f}_{i \sigma}^{\dagger}\rangle\rangle\approx\langle\delta n_{r\overline{\sigma'}}\tilde{f}_{r \sigma'}^{\dagger}\tilde{f}_{l \sigma'}\rangle \langle\langle \tilde{f}_{j \sigma}; \tilde{f}_{i \sigma}^{\dagger}\rangle\rangle
\end{equation*}
in the case of Eq.~\eqref{eq:eom_fff1}, and
\begin{equation*}
    \langle\langle \tilde{f}_{l \sigma'}^{\dagger}\delta n_{r\overline{\sigma'}}\tilde{f}_{r \sigma'}\tilde{f}_{j \sigma}; \tilde{f}_{i \sigma}^{\dagger}\rangle\rangle\approx\langle\tilde{f}_{l \sigma'}^{\dagger}\delta n_{r\overline{\sigma'}}\tilde{f}_{r \sigma'}\rangle \langle\langle \tilde{f}_{j \sigma}; \tilde{f}_{i \sigma}^{\dagger}\rangle\rangle
\end{equation*}   
%
in the case of Eq.~\eqref{eq:eom_fff2}. Thus, when adding Eqs.~\eqref{eq:eom_fff1} and~\eqref{eq:eom_fff2}, the respective lines cancel  
and the result reads
\begin{equation}\label{eq:eom_fff3}
    \begin{split}
        &i\partial_t\langle\langle \big(\tilde{f}_{j \sigma'}^{\dagger}\tilde{f}_{l \sigma'}+\tilde{f}_{l \sigma'}^{\dagger}\tilde{f}_{j \sigma'}\big)\tilde{f}_{j \sigma}; \tilde{f}_{i \sigma}^{\dagger}\rangle\rangle^c \\
        &=
        \frac{U}{2}\lambda_{lj}\langle\langle \tilde{f}_{j \sigma}; \tilde{f}_{i \sigma}^{\dagger}\rangle\rangle\\
        &+\ U\sum_{r\neq j}\lambda_{lr} \langle\langle  \delta n_{r \overline{\sigma'}}\big(\tilde{f}_{j \sigma'}^{\dagger}\tilde{f}_{r \sigma'}-\tilde{f}_{r \sigma'}^{\dagger}\tilde{f}_{j \sigma'}\big)\tilde{f}_{j \sigma}; \tilde{f}_{i \sigma}^{\dagger}\rangle\rangle.
    \end{split}
\end{equation}
Equations~\eqref{eq:eom_fpsif-psiff} and~\eqref{eq:eom_fff3} have the same structure as Eqs.~\eqref{eq:eom_nnf_approx} and~\eqref{eq:eom_nf}, so in direct analogy with Eq.~\eqref{eq:nnf_freq} we find
\begin{equation}\label{eq:gf_psiff}
    \begin{split}
        &\langle\langle \delta n_{l \overline{\sigma'}}\big(\tilde{f}_{j \sigma'}^{\dagger}\tilde{f}_{l \sigma'}-\tilde{f}_{l \sigma'}^{\dagger}\tilde{f}_{j \sigma'}\big)\tilde{f}_{j \sigma}; \tilde{f}_{i \sigma}^{\dagger}\rangle\rangle_{\omega} \\
    =& \frac{2\omega}{U}\sum_{r\neq j}W_{lr, j}(\omega)t_{rj}^{\scriptscriptstyle U}(\omega)\langle\langle\psi_{j \sigma}; \tilde{f}_{i \sigma}^{\dagger}\rangle\rangle_{\omega}.
    \end{split}
\end{equation}
Plugging this result back into Eq.~\eqref{eq:resonance_broadening}, we obtain the resonance broadening correction for $N=2$,
\begin{align}
\label{eq:resonance_broadening_correction}
\begin{split}
    \langle\langle [\delta n_{j\bar{\sigma}},\tilde{H}]\tilde{f}_{l \sigma}; \tilde{f}_{i  \sigma}^{\dagger}\rangle\rangle= N c(\omega)\omega\langle\langle \psi_{j \sigma}; \tilde{f}_{i \sigma}^{\dagger}\rangle\rangle_{\omega}.
\end{split}
\end{align}
In Appendix~\ref{sec:evenN}, we show that this equation holds for all (even) $N$. 

Substituting Eqs.~\eqref{eq:scattering_correction} and~\eqref{eq:resonance_broadening_correction} into the equation of motion for the trion, Eq.~\eqref{eq:eom_psi}, yields a closed set of equations together with Eq.~\eqref{eq:eom_f}. From these we derive the self-energy of the flat bands,
\begin{equation}\label{eq:self-energy_HubbardIII}
    \Sigma_{\rm III}(\bm k, \omega)= \frac{U^2}{4\omega}\frac{k^2}{k^2+k_{\star}^2}\!\left[1-(N+1) s^2\ln\left(1-\frac{U^2}{4\omega^2}\right)\!\right].
\end{equation}

\section[Finite M]{Finite $M$}\label{sec:M}
In the following, we compute the spectra and quasiparticle relaxation rates in the presence of a small $M\ll U$.
The corresponding single-particle Hamiltonian $\tilde{H}_0=PH_0P$ generates intersite hopping and hybridizes $\alpha=1,2$ orbitals [see Eq.~\eqref{eq:PH0P}],
\begin{align}
    [\tilde{f}_{i\sigma},\tilde{H}_0] = \sum_{j}\sum_{\mu}t_{ij}^{\sigma\mu} \tilde{f}_{j\mu},
\end{align}
where $t_{ij}^{\sigma\mu}\equiv \sum_{\bm k}h({\bm k})_{\sigma,\mu}e^{i\bm k\cdot(\bm R_i-\bm R_j)}/N_s$. For the Hamiltonian given by Eq.~\eqref{eq:h}, the on-site potential vanishes, $t_{ii}^{\sigma\mu}=0$, and the intersite hopping $|t_{ij}^{\sigma\mu}|\sim O(s^2) (i\neq j)$ .

The equations of motion for $\tilde{f}_{i \sigma}$ and $\psi_{i \sigma}$, Eqs.~\eqref{eq:eom_f} and~\eqref{eq:eom_psi}, should be generalized to
\begin{align}
    i\partial_t \langle\langle \tilde{f}_{j \sigma}; \tilde{f}_{i \rho}^{\dagger}\rangle\rangle &= \lambda_{ji}\delta_{\sigma,\rho}\delta(t) + U \sum_{l} \lambda_{jl}\langle\langle \psi_{l\sigma}; \tilde{f}_{i \rho}^{\dagger}\rangle\rangle\notag\\
    &\quad\   +  \sum_{l}\sum_{\mu}t_{jl}^{\sigma\mu} \langle\langle \tilde{f}_{l\mu}; \tilde{f}_{i \rho}^{\dagger}\rangle\rangle  ,\label{eq:eom_f_M}\\
    i\partial_t \langle\langle \psi_{j\sigma}; \tilde{f}_{i \rho}^{\dagger}\rangle\rangle &= \lambda_{ji}[\langle\delta n_{j\bar{\sigma}}\rangle\delta_{\sigma,\rho} + \langle \tilde{f}_{j\rho}^{\dagger}\tilde{f}_{j\sigma}\rangle(1-\delta_{\sigma,\rho})]\delta(t)\notag\\
    &\quad\ +U\langle\delta n_{j\sigma}^2\rangle\langle\langle \tilde{f}_{j \sigma}; \tilde{f}_{i \rho}^{\dagger}\rangle\rangle \notag\\
    &\quad\ + U \sum_{l\neq j} \lambda_{lj}\langle\langle \delta n_{j \bar{\sigma}}\delta n_{l \sigma}\tilde{f}_{l\sigma}; \tilde{f}_{i \rho}^{\dagger}\rangle\rangle\notag\\
    &\quad\ + \sum_{l\neq j}\sum_{\mu} t_{jl}^{\sigma\mu} \langle\langle  \delta n_{j \bar{\sigma}}\tilde{f}_{l \mu}; \tilde{f}_{i \rho}^{\dagger}\rangle\rangle\notag\\
    &\quad\ + \langle\langle[\delta n_{j\bar{\sigma}}, \tilde{H}_0+\tilde{H}_U] \tilde{f}_{j\sigma}, \tilde{f}_{i\rho}^{\dagger}\rangle\rangle. \label{eq:eom_psi_M}
\end{align}
On the right-hand side of Eq.~\eqref{eq:eom_psi_M}, the first line vanishes in the paramagnetic Mott state. We note that we used $\langle \tilde{f}_{j\rho}^{\dagger}\tilde{f}_{j\sigma}\rangle=\delta_{\rho,\sigma}$ because of spin-valley conservation as well as the different $C_{3z}$ eigenvalues associated with $\alpha=1,2$ orbitals in each spin-valley sector,
\begin{equation}
 C_{3z}^{-1}\tilde{f}_{i \eta \alpha}C_{3z} = e^{\frac{2\pi i \alpha}{3}}\tilde{f}_{i \eta\alpha},   
\end{equation}
required by the $C_{3}$ invariance of $H_0$ [see Eq.~\eqref{eq:H_0_full}].
To leading order in $s^2$, we can drop all terms with $l\neq j$ on the right-hand side of Eq.~\eqref{eq:eom_psi_M}, and solve the closed equations. The resulting Hubbard-I approximation for the flat-band single-particle Green's function reads
%
\begin{equation}\label{eq:g_M}
    g(\bm k,\omega)_{\sigma,\rho} = \left[\Big(\omega - \frac{U^2}{4\omega}\frac{k^2}{k^2+k_{\star}^2}\Big)\mathcal{I}- h(\bm k)\right]^{-1}_{\sigma,\rho},
\end{equation}
where $\mathcal{I}$ denotes the identity matrix in flavor space. Interestingly, this Green's function matrix can be diagonalized in the same basis as the single-particle Hamiltonian $h(\bm k)$ [see Eq.~\eqref{eq:h}],
\begin{align}
    d_{\bm k\eta\pm} = \frac{e^{i\theta_{\bm k}}}{\sqrt{2}}d_{\bm k\eta 1} \pm \frac{e^{-i\theta_{\bm k}}}{\sqrt{2}}d_{\bm k\eta 2}, \label{eq:basis_band}
\end{align}
despite the fact that the interacting Hamiltonian can induce interband scattering. The resulting quasiparticle dispersion is given by Eq.~\eqref{eq:fourmodes}.

Next, we compute the Green's functions in Eq.~\eqref{eq:eom_psi_M} to the next order in $s^2$. The third and fourth lines correspond to the scattering correction. The corresponding equations of motion for the Green's functions follow as a straightforward generalization of Eqs.~\eqref{eq:eom_nnf} and~\eqref{eq:eom_nf},
\begin{align}
    &i\partial_t \langle\langle \delta n_{j\bar{\sigma}}\delta n_{l\bar{\sigma}}\tilde{f}_{l\sigma}; \tilde{f}_{i \rho}^{\dagger}\rangle\rangle = U\langle\delta n_{j\bar{\sigma}}^2\rangle\langle\langle \delta n_{j\bar{\sigma}}\tilde{f}_{l\sigma}; \tilde{f}_{i \rho}^{\dagger}\rangle\rangle,\\
    &i\partial_t \langle\langle \delta n_{j\bar{\sigma}}\tilde{f}_{l\sigma}; \tilde{f}_{i \rho}^{\dagger}\rangle\rangle = U\sum_{r}\lambda_{lr} \langle\langle \delta n_{j\bar{\sigma}}\delta n_{r\bar{\sigma}}\tilde{f}_{r\sigma}; \tilde{f}_{i \rho}^{\dagger}\rangle\rangle \notag\\
    &\qquad\qquad\qquad\qquad\quad + \sum_{r}\sum_{\mu} t_{jr}^{\sigma\mu} \langle\langle  \delta n_{j \bar{\sigma}}\tilde{f}_{r \mu}; \tilde{f}_{i \rho}^{\dagger}\rangle\rangle.
\end{align}
The scattering correction in Eq.~\eqref{eq:eom_psi_M} becomes
\begin{align}\label{eq:scattering_correction_M}
    &U\sum_{l\neq j}\lambda_{jl} \langle\langle \delta n_{j \bar{\sigma}}\delta n_{l\bar{\sigma}}\tilde{f}_{l \sigma}; \tilde{f}_{i \rho}^{\dagger}\rangle\rangle_{\omega}\notag\\
    &+\sum_{l\neq j}\sum_{\mu}t_{jl}^{\sigma\mu}\langle\langle \delta n_{j \bar{\sigma}}\tilde{f}_{l \mu}; \tilde{f}_{i \rho}^{\dagger}\rangle\rangle_{\omega} = c(\omega) \omega \langle\langle \tilde{\psi}_{j \sigma}; \tilde{f}_{i \rho}^{\dagger}\rangle\rangle_{\omega}, 
\end{align}
with a flavor-independent coefficient that generalizes Eq.~\eqref{eq:c1_def},
\begin{align}
    c(\omega) &= \frac{\sum_{\bm k} \text{tr}\left[ g(\bm k,\omega)\left(\omega\bar{\lambda}(\bm k)\mathcal{I}+h(\bm k)\right)\right]}{\omega\sum_{\bm k}\text{tr}\left[ g(\bm k,\omega)\right]},
    \label{eq:c_M}
\end{align}
%
%
We have also checked that when the single-particle Hamiltonian Eq.~\eqref{eq:h} is included, the resonance broadening correction is identical to Eq.~\eqref{eq:resonance_broadening_correction}, with the coefficient $c(\omega)$ given by Eq.~\eqref{eq:c_M}.
Substituting that and Eq.~\eqref{eq:scattering_correction_M} into Eq.~\eqref{eq:eom_psi_M}, we obtain the self-energy
\begin{equation}\label{eq:self_energy_M}
    \Sigma_{\text{III}}(\bm k, \omega) = \frac{U^2}{4\omega [1-(N+1)c(\omega)]}\frac{k^2}{k^2+k_{\star}^2}.
\end{equation}
To first order in $s^2$, the real and imaginary parts of the self-energy read
\begin{subequations}\label{eq:self-energy_HubbardIIIM}
\begin{equation}\label{eq:self-energy_re_HubbardIIIM}
\begin{split}
\mathrm{Re}\,\Sigma_{\mathrm{III}}(\bm k,\omega)
&= \frac{U^2}{4\omega}\frac{k^2}{k^2+k_{\star}^2}
\bigg\{ 1 -\frac{N+1}{2}s^2 \bigg[ \\
&\quad\ \ln\left|\frac{\omega^2-U^2/4}{\omega(\omega-M)}\right|
+ \ln\left|\frac{\omega^2-U^2/4}{\omega(\omega+M)}\right| \\
&\quad\ + \frac{M}{\omega}\ln\left|\frac{\omega-M}{\omega+M}\right|
\bigg] \bigg\}
\end{split}
\end{equation}
and 
\begin{equation}\label{eq:self-energy_im_HubbardIIIM}
\begin{split}
    \text{Im}\Sigma_{\text{III}}(\bm k,\omega) &= -(N+1)\pi s^2\frac{U^2}{4\omega}\frac{k^2}{k^2+k_{\star}^2}\\
    &\ \ \ \times\left[\Theta(\frac{U}{2}-|\omega|) + \frac{M-|\omega|}{2|\omega|}\Theta(M-|\omega|)\right],
\end{split}
\end{equation}
\end{subequations}
respectively. The second term in the bracket diverges as $\omega\rightarrow 0$, suggesting that the first-order expansion in $s^2$  breaks down within a narrow energy range $|\omega|\ll (N+1)\pi s^2M$. This energy scale also corresponds to the trion relaxation rate in the limit of small $k$, see Eq.~\eqref{eq:relaxation_trion}.

\section{Conclusion and Discussion}

In this work, we studied the electron spectra of the paramagnetic Mott state in charge neutral magic-angle twisted bilayer graphene (MATBG). Despite the topological obstruction to constructing exponentially localized symmetric Wannier orbitals from MATBG flat bands, we show that the dominant electron-electron interactions within the flat bands can be described by an on-site Hubbard interaction in a nonorthogonal basis. The intersite orbital overlap is set by the fraction $s^2$ of the moiré Brillouin zone over which $f$- and $c$-electrons strongly hybridize. Treating $s^2$ as a small parameter, we compute the electron self-energy to the first order in $s^2$ using the equation of motion method, and obtain concrete predictions for the electron spectral function. These can be probed by measurements with a quantum twisting microscope, which allow for spectroscopy with a high resolution of energy and momentum \cite{inbar2023quantum,xiao2025interacting,wei2025dirac,wei2025theory}.


Our approach is adapted from the Hubbard-III approximation for the Hubbard model. We highlight the similarities and differences in the resulting self-energy. In both cases, the Hubbard-III self-energy is diagonal in the site basis, $\Sigma_{ij}(\omega)\propto \delta_{ij}$, but the nonorthogonality of the orbitals renders the self-energy momentum dependent in our case. Moreover, we find well-defined quasiparticles with relaxation rate much smaller than the bandwidth ($\sim U$) near the $\Gamma$ point where Berry curvature concentrates. This behavior contrasts sharply with the conventional strong-coupling Hubbard model (hopping $t\ll U$), where the bandwidth $W\sim t$ and the quasiparticle broadening$\sim t^2/W\sim W$ are of the same order.

While the first-order expansion in $s^2$ captures qualitative features of the electron spectra, higher-order corrections are expected to be important in several respects. First, the self-energy in Eq.~\eqref{eq:self_energy_M} exhibits a logarithmic divergence at $\omega=0,\pm M,\pm U/2$, signaling the breakdown of the first-order expansion near the band edges. 
Second, 
while the experimental value of $s^2$ remains difficult to determine at present, some theoretical estimates such as $s\sim 0.25$ in Ref.~\cite{ledwidth2025nonlocal} suggest $(N+1)s^2\sim 1$. This suggests that  higher-order corrections can be quantitatively significant in real MATBG samples.  Moreover, our theory is restricted to the paramagnetic Mott state, which exists within a finite temperature window $s^2U\ll k_BT\ll U$ \cite{ledwidth2025nonlocal,hu2025projected}. At lower temperatures, nonlocal exchange interactions can drive flavor polarization \cite{song2022magic,xie2020nature,bultinck2020ground,zhang2020correlated,lian2021twisted}, accompanied by strong nonlocal collective fluctuations beyond the scope of Hubbard-III approximation. 
Extending the present framework to incorporate these effects remains an important open direction. The modified Hubbard model formulated here can serve as a useful starting point for more advanced approaches, such as dynamical mean-field theory and its cluster extensions \cite{georges1996dynamical}.

We have focused on the dominant on-site repulsion among $f$-electrons. Additional interactions, including those between $c$-and $f$-electrons and among $c$-electrons, can further reduce the quasiparticle lifetime. However, due to the low density of states in the Hubbard-I/III spectra near the $\Gamma$ point, these residual interaction effects are expected to be weak in this region of the moir\'e Brillouin zone, as proposed in Ref.~\cite{hu2026thf}.

Although our analysis focuses on half filling of the flat bands ($\nu=0$), the method can be extended to other integer fillings. It will also be interesting to compute thermodynamic and transport properties in future work \footnote{P. Nosov, E. Khalaf, and P.J. Ledwith (2026), to appear}. 
More broadly, our approach may be applicable to other topological flat-band systems that admit a heavy-fermion description \cite{liu2026ideal,xie2025superconductivity}, providing a route to understand correlation effects in flat bands with nontrivial quantum geometry.


\begin{acknowledgments}
    We thank Erez Berg, Andrei Bernevig, Dumitru Călugăru, Antoine Georges, Shahal Ilani, Patrick Ledwith, and Yaar Vituri for helpful discussions. N.W.\ acknowledges support through the Yale Prize Postdoctoral Fellowship in Condensed Matter Theory. Research at Yale was supported by NSF Grant No.\ DMR-2410182 and by the 	Air Force Office of Scientific Research under Award No.\ FA95502510287. Research at Freie Universit\"at Berlin was supported by Deutsche Forschungsgemeinschaft through CRC 183 (project C02), CRC 1772 (project B06), as well as the German Excellence Strategy - EXC3112/1 - 533767171 (Center for Chiral Electronics).
\end{acknowledgments}

\clearpage
\newpage

\appendix

\section{Hubbard-I Green's function}\label{sec:gf}
In this section, we evaluate the Hubbard-I approximation for the electron Green's functions at charge neutrality, and then use them to compute equal-time correlation functions. We first write the equations of motion, Eqs.~\eqref{eq:eom_f_M} and~\eqref{eq:eom_psi_M}, truncated at the Hubbard-I level and Fourier transformed to frequency space,
\begin{align}
   \omega  \langle\langle \tilde{f}_{j \sigma}; \tilde{f}_{i \rho}^{\dagger}\rangle\rangle_{\omega} &= \lambda_{ji}\delta_{\sigma,\rho} + \sum_{l}\lambda_{jl}\langle\langle \psi_{l \sigma}; \tilde{f}_{i \rho}^{\dagger}\rangle\rangle_{\omega}\notag\\
   &\quad +  \sum_{j}\sum_{\mu}t_{ji}^{\sigma\mu} \langle\langle \tilde{f}_{i\mu}; \tilde{f}_{i \rho}^{\dagger}\rangle\rangle_{\omega},  \\
   \omega  \langle\langle \psi_{j \sigma}; \tilde{f}_{i \rho}^{\dagger}\rangle\rangle_{\omega} &= U\langle\delta n_{j\bar{\sigma}}^2\rangle \langle\langle \tilde{f}_{j\sigma}; \tilde{f}_{i \rho}^{\dagger}\rangle\rangle_{\omega}.\label{eq:eom_psi_freq}
\end{align}
From these two closed equations, we obtain
\begin{subequations}
    \begin{equation}\label{eq:g_ff}
        \langle\langle \tilde{f}_{j \sigma}; \tilde{f}_{i \rho}^{\dagger}\rangle\rangle_{\omega} = \int_{\bm k} g(\bm k,\omega)_{\sigma,\rho} \hat{\lambda}(\bm k) e^{i\bm k\cdot (\bm R_j-\bm R_i)},
    \end{equation}
    \begin{equation}\label{eq:g_psif}
        \langle\langle \psi_{j \sigma}; \tilde{f}_{i \rho}^{\dagger}\rangle\rangle_{\omega} = \frac{U\langle\delta n_{j\bar{\sigma}}^2\rangle}{\omega}\int_{\bm k} g(\bm k,\omega)_{\sigma,\rho} \hat{\lambda}(\bm k) e^{i\bm k\cdot (\bm R_j-\bm R_i)},\\
    \end{equation}
\end{subequations}
%
where $\int_{\bm k} =\int_{\text{mBZ}} d^2 k/\Omega_{m}$. The expression for the retarded Green's function $g(\bm k,\omega)$ is shown in Eq.~\eqref{eq:g_M}. In particular, its diagonal element reads
\begin{equation}\label{eq:g_ff_diagonal}
     g(\bm k,\omega)_{\sigma,\sigma}= \frac{1}{2}\sum_{\beta=\pm} \frac{\omega_{+}}{\omega_{+}(\omega_{+}+\beta|\hat{t}_{\bm k}|)-\frac{U^2}{4}\hat{\lambda}(\bm k)}.
\end{equation}
Here, $\omega_{+} = \omega + i0^{+}$, and we used $\langle\delta n_{j\bar{\sigma}}^2\rangle= 1/4$ at low temperatures, $k_BT\ll U$. We also introduced the Fourier transform of $\lambda_{ij}$ in momentum space,
\begin{equation}
    \hat{\lambda}(\bm k) = \sum_{i}\lambda_{ij}e^{-i\bm k\cdot(\bm R_i-\bm R_j)} = \frac{1}{z}\frac{k^2}{k^2+ k_{\star}^2}.
\end{equation}

According to the equilibrium KMS condition, the equal-time correlation function of two operators $A$ and $B$ (with odd fermion parity) is related to their retarded and advanced Green's functions,
\begin{equation}\label{eq:spectral_theorem}
    \langle BA\rangle = i\int_{-\infty}^{\infty} \frac{d\omega}{2\pi} n_F(\omega)\left[\langle\langle A; B\rangle\rangle_{\omega} - \langle\langle A; B\rangle\rangle_{\omega}^{-}\right],
\end{equation}
where $n_F(\omega)$ denotes the Fermi-Dirac distribution function and the notation $\langle\langle A; B\rangle\rangle_{\omega}^{-}$ represents the advanced Green's function, which can be obtained via the substitution $\omega_{+}= \omega + i0^{+}\rightarrow \omega + i0^{-}$ in $\langle\langle A; B\rangle\rangle_{\omega}$. 

Plugging Eqs.~\eqref{eq:g_ff} and~\eqref{eq:g_ff_diagonal} into Eq.~\eqref{eq:spectral_theorem}, we obtain the following equal-time correlation function of the Mott semimetal at charge neutrality,
\begin{equation}\label{eq:corr_ff}
    \langle \tilde{f}_{i\sigma}^{\dagger} \tilde{f}_{j\sigma}\rangle = \frac{1}{2}\lambda_{ij}.
\end{equation}
This indicates $\langle \tilde{f}_{i\sigma}^{\dagger} \tilde{f}_{j\sigma}\rangle= \langle \tilde{f}_{j\sigma}^{\dagger} \tilde{f}_{i\sigma}\rangle$.

Next, we insert Eqs.~\eqref{eq:g_psif} and~\eqref{eq:g_ff_diagonal} into Eq.~\eqref{eq:spectral_theorem}, and use $n_{F}(\omega)\approx \theta(-\omega)$ at low temperatures ($k_BT\ll U$) to simplify the integral. The result is 
\begin{equation}\label{eq:corr_fpsi}
    \langle \tilde{f}_{i\sigma}^{\dagger} \psi_{j\sigma}\rangle = -\frac{1}{4}\xi_{ij},
\end{equation}
where we introduced
\begin{equation}
    \xi_{ji}=\int_{\bm k}\frac{U\hat{\lambda}(\bm k)}{\sqrt{|\hat{t}_{\bm k}|^2+ U^2\hat{\lambda}(\bm k)}}e^{i\bm k\cdot(\bm R_j-\bm R_i)}.
\end{equation}
From Eq.~\eqref{eq:corr_fpsi}, we verify that $\langle \psi_{i\sigma}^{\dagger} \tilde{f}_{j\sigma}\rangle=\langle \tilde{f}_{j\sigma}^{\dagger} \psi_{i\sigma}\rangle^*=\langle \tilde{f}_{i\sigma}^{\dagger} \psi_{j\sigma}\rangle$.

\section[W]{$W_{jl,i}$}\label{sec:W}
In this section, we discuss the solution of a class of linear equations that frequently arise in the Hubbard-III approximation. These equations take the following form,
\begin{equation}\label{eq:eom_impurity}
    \omega {\bm G}_{lj}(\omega) = \sum_{r\neq j} \hat{\mathcal{H}}_{lr}(\omega) {\bm G}_{rj}(\omega) + {\bm S}_{lj}(\omega),\quad l\neq j,
\end{equation}
where ${\bm G}_{lj}$ denotes a Green's function. Note that ${\bm G}$ can in general depend on multiple real-space coordinates, and we suppress those that are irrelevant for this discussion. The bold symbol indicates that the Green's functions are allowed to be a vector with multiple components (flavors).
This equation effectively describes a particle hopping on a lattice that contains a single impurity at site $j$ but is otherwise uniform. $\hat{\mathcal{H}}_{lr}$ represents a translationally invariant `hopping' matrix. The impurity hybridizes with the particle and generates a source term ${\bm S}(\omega)$ in the equation.

Because this equation does not involve $\hat{\bm G}_{jj}$ and ${\bm S}_{jj}$, we can set $\hat{\bm G}_{jj}={\bm S}_{jj}=0$ at the moment and rewrite the equation as
\begin{equation}\label{eq:eom_impurity2}
\begin{split}
    \omega {\bm G}_{lj}(\omega) = &\sum_{r} \hat{\mathcal{H}}_{lr}(\omega) {\bm G}_{rj}(\omega) + {\bm S}_{lj}(\omega) \\
    &- \delta_{lj}\sum_{r}\hat{\mathcal{H}}_{jr}(\omega) {\bm G}_{rj}(\omega),\ \ \ \forall l.
\end{split}
\end{equation}
For $l\neq j$, this equation is equivalent to Eq.~\eqref{eq:eom_impurity}, while for $l=j$, both sides of the equation vanish. 
Therefore,
\begin{equation}\label{eq:G}
    {\bm G}_{lj}(\omega) = \sum_{r} \hat{g}_{lr}(\omega)  {\bm S}_{rj}(\omega) - \hat{g}_{lj}(\omega)\sum_{r}\hat{\mathcal{H}}_{jr}(\omega) {\bm G}_{rj}(\omega),
\end{equation}
with
\begin{equation}
    \hat{g}_{lj}(\omega) = \big[\omega\hat{\mathcal{I}} - \hat{\mathcal{H}}(\omega)\big]_{lj}^{-1}.
\end{equation}

This leads to the self-consistent equation
\begin{align}
    \bm C(\omega)&\equiv \sum_{r}\hat{\mathcal{H}}_{jr}(\omega){\bm G}_{rj}(\omega) \notag\\
    &= \sum_{lr} \hat{\mathcal{H}}_{jl}(\omega)\hat{g}_{lr}(\omega) {\bm S}_{rj}(\omega) - \sum_{l}\hat{\mathcal{H}}_{jl}(\omega)\hat{g}_{lj}(\omega)\bm C(\omega)\notag\\
    &= \hat{g}_{jj}(\omega)^{-1}\sum_{l}\hat{g}_{jl}(\omega)S_{lj}(\omega).\label{eq:C}
\end{align}
To arrive at the last equation, we have used $\sum_{l}\hat{\mathcal{H}}_{jl}(\omega)\hat{g}_{lr}(\omega)= \omega\hat{g}_{jr}(\omega)-\delta_{jr}$ and $S_{jj}=0$. Plugging Eq.~\eqref{eq:C} into Eq.~\eqref{eq:G}, we obtain the solution to Eq.~\eqref{eq:eom_impurity}, the main result of this section,
\begin{align}
    &{\bm G}_{lj}(\omega) = \sum_{r \neq j} \hat{W}_{lr, j}(\omega) {\bm S}_{rj}(\omega),\\
    &\hat{W}_{lr, j}(\omega) = \hat{g}_{jl}(\omega) - \hat{g}_{lj}(\omega)\hat{g}_{jj}(\omega)^{-1}\hat{g}_{jr}(\omega).\label{eq:W_matrix}
\end{align}
If $\hat{g}(\omega)$ is proportional to an identity matrix in flavor space, Eqs.~\eqref{eq:W_matrix} and~\eqref{eq:C} reduce to Eqs.~\eqref{eq:W} and~\eqref{eq:c1_def} in the main text, respectively. 

\section[Resonance broadening correction for general even N]{Resonance broadening corrections for general even $N$}\label{sec:evenN}

The original work introducing the Hubbard-III approximation considers the Hubbard model with two spin flavors \cite{hubbard1964III}. To our knowledge, the extension to flavor number $N>2$ has not been discussed. Here we present the calculation for general even $N$. For simplicity, we focus on the chiral-flat limit, $M=0$ below.

The derivation of the scattering correction [Eq.~\eqref{eq:scattering_correction}] in Sec.~\ref{sec:scattering_correction} is applicable to general even $N$. For resonance broadening corrections, compared to the $N=2$ case in Sec.~\ref{sec:resonance}, we find additional terms on the right-hand sides of the equations of motion in Eqs.~\eqref{eq:eom_fpsif}, ~\eqref{eq:eom_psiff}, ~\eqref{eq:eom_fff1}, and~\eqref{eq:eom_fff2}. These additional terms are summarized below. To avoid repetition, we use ellipses to represent terms that have already appeared on the right-hand side of these equations.

(i) Eq.~\eqref{eq:eom_fpsif} remains unchanged;

(ii) Generalization of Eq.~\eqref{eq:eom_fff1}:
\begin{equation}
    \begin{split}
        &i\partial_t\langle\langle \tilde{f}_{j \sigma'}^{\dagger}\tilde{f}_{l \sigma'}\tilde{f}_{j \sigma}; \tilde{f}_{i \sigma}^{\dagger}\rangle\rangle^{c} \\
        =\ & U\lambda_{lj} \langle\langle (\tilde{f}_{j \sigma'}^{\dagger}\tilde{f}_{j \sigma'}-\frac{1}{2})\delta n_{j \overline{\sigma\sigma'}}\tilde{f}_{j \sigma}; \tilde{f}_{i \sigma}^{\dagger}\rangle\rangle +...;\label{eq:eom_fff1_N}
    \end{split}
\end{equation}

(iii) Generalization of Eq.~\eqref{eq:eom_psiff}:
\begin{align}
        & i\partial_t\langle\langle \delta n_{l \overline{\sigma'}}\tilde{f}_{l \sigma'}^{\dagger}\tilde{f}_{j \sigma'}\tilde{f}_{j \sigma}; \tilde{f}_{i \sigma}^{\dagger}\rangle\rangle  \notag\\
        = &\ 2U \langle\langle \delta n_{l \overline{\sigma'}}\tilde{f}_{l \sigma'}^{\dagger}\delta n_{j \overline{\sigma\sigma'}}\tilde{f}_{j \sigma'}\tilde{f}_{j \sigma}; \tilde{f}_{i \sigma}^{\dagger}\rangle\rangle +...;
        \label{eq:eom_psiff_N}
\end{align}

(iv) Generalization of Eq.~\eqref{eq:eom_fff2}:
\begin{align}
    &i\partial_t\langle\langle \tilde{f}_{l \sigma'}^{\dagger}\tilde{f}_{j \sigma'}\tilde{f}_{j \sigma}; \tilde{f}_{i \sigma}^{\dagger}\rangle\rangle^{c} \notag\\
    \approx&\ -U\lambda_{lj} \langle\langle (\tilde{f}_{j \sigma'}^{\dagger}\tilde{f}_{j \sigma'}-\frac{1}{2})\delta n_{j \overline{\sigma\sigma'}}\tilde{f}_{j \sigma}; \tilde{f}_{i \sigma}^{\dagger}\rangle\rangle \notag \\
    &+ 2U \langle\langle (\tilde{f}_{l \sigma'}^{\dagger}\tilde{f}_{j \sigma'}-\langle\tilde{f}_{l \sigma'}^{\dagger}\tilde{f}_{j \sigma'}\rangle)\delta n_{j \overline{\sigma\sigma'}}\tilde{f}_{j \sigma}; \tilde{f}_{i \sigma}^{\dagger}\rangle\rangle+....\label{eq:eom_fff2_N}
\end{align}

Here we introduced the notation
\begin{equation}
    \delta n_{i \overline{\sigma\sigma'}}\equiv \sum_{\sigma''\neq \sigma,\sigma'}\left(\tilde{f}_{i \sigma'}^{\dagger}\tilde{f}_{i\sigma'}-\frac{1}{2}\right).
\end{equation}
We solve these equations of motions using a slightly different approach compared to the main text. First, we Fourier transform the four equations in (i)-(iv) to frequency space. We then plug Eqs.~\eqref{eq:eom_fff1_N} and~\eqref{eq:eom_fff2_N} into Eq.~\eqref{eq:eom_fpsif} and~\eqref{eq:eom_psiff_N}, respectively. We arrive at

\begin{equation}\label{eq:eom_fpsif_freq_N}
    \begin{split}
        &\omega \langle\langle \tilde{f}_{j \sigma'}^{\dagger}\delta n_{l \overline{\sigma'}}\tilde{f}_{l \sigma'}\tilde{f}_{j \sigma}; \tilde{f}_{i \sigma}^{\dagger}\rangle\rangle_{\omega} \\
        &-\sum_{r \neq j}t_{rj}^{\scriptscriptstyle{U}}(\omega)\langle\langle \tilde{f}_{j \sigma'}^{\dagger}\delta n_{r \overline{\sigma'}}\tilde{f}_{r \sigma'}\tilde{f}_{j \sigma}; \tilde{f}_{i \sigma}^{\dagger}\rangle\rangle_{\omega} =S_{l,ji}^{(1)}(\omega),
    \end{split}
\end{equation}
\begin{equation}\label{eq:eom_psiff_freq_N}
    \begin{split}
        &\omega \langle\langle \delta n_{l\overline{\sigma'}}\tilde{f}_{l\sigma'}^{\dagger}\tilde{f}_{j\sigma'}\tilde{f}_{j\sigma}; \tilde{f}_{i\sigma}^{\dagger}\rangle\rangle_{\omega} \\
        &-\sum_{r\neq j}t_{rl}^{\scriptscriptstyle{U}}(\omega) \langle\langle \delta n_{r\overline{\sigma'}}\tilde{f}_{r\sigma'}^{\dagger}\tilde{f}_{j\sigma'}\tilde{f}_{j\sigma}; \tilde{f}_{i\sigma}^{\dagger}\rangle\rangle_{\omega}=S_{l,ji}^{(2)}(\omega),
    \end{split}
\end{equation}
with the effective hopping amplitude $t_{rj}^{\scriptscriptstyle{U}}(\omega)=\frac{U^2}{4\omega}\lambda_{rj}$ and
\begin{equation}\label{eq:S1}
    \begin{split}
        S_{l,ji}^{(1)}(\omega)=& \frac{\omega}{U}t_{lj}^{\scriptscriptstyle{U}}(\omega)\langle\langle \psi_{j \sigma}; \tilde{f}_{i \sigma}^{\dagger}\rangle\rangle_{\omega}\\
        & + t_{jl}^{\scriptscriptstyle{U}}(\omega) \langle\langle (\tilde{f}_{j \sigma'}^{\dagger}\tilde{f}_{j \sigma'}-\frac{1}{2})\delta n_{j \overline{\sigma\sigma'}}\tilde{f}_{j \sigma}; \tilde{f}_{i \sigma}^{\dagger}\rangle\rangle_{\omega}\\
        &+\langle\tilde{f}_{j \sigma'}^{\dagger} \psi_{l \sigma'}\rangle \big(\lambda_{ji}+U\sum_{r\neq j}\lambda_{jr}\langle\langle \psi_{r \sigma}; \tilde{f}_{i \sigma}^{\dagger}\rangle\rangle_{\omega}\big)\\
        &-  \sum_{r\neq j}t_{rj}^{\scriptscriptstyle{U}}(\omega)\langle\psi_{r \sigma'}^{\dagger}\tilde{f}_{l \sigma'}\rangle \langle\langle\tilde{f}_{j \sigma}; \tilde{f}_{i \sigma}^{\dagger}\rangle\rangle_{\omega},
    \end{split}
\end{equation}
\begin{equation}\label{eq:S2}
    \begin{split}
        S_{l,ji}^{(2)}(\omega)=&-\frac{\omega}{U}t_{lj}^{\scriptscriptstyle{U}}(\omega)\langle\langle\psi_{j\sigma}; \tilde{f}_{i\sigma}^{\dagger}\rangle\rangle_{\omega} \\
        &- \frac{U^2}{2\omega} \langle\langle (\tilde{f}_{l\sigma'}^{\dagger}\tilde{f}_{j\sigma'}-\langle\tilde{f}_{l \sigma'}^{\dagger}\tilde{f}_{j \sigma'}\rangle)\delta n_{j\overline{\sigma\sigma'}}\tilde{f}_{j\sigma}; \tilde{f}_{i\sigma}^{\dagger}\rangle\rangle_{\omega}\\
        &+2U \langle\langle \delta n_{l\overline{\sigma'}}\tilde{f}_{l\sigma'}^{\dagger}\delta n_{j\overline{\sigma\sigma'}}\tilde{f}_{j\sigma'}\tilde{f}_{j\sigma}; \tilde{f}_{i\sigma}^{\dagger}\rangle\rangle_{\omega}\\
        &+ t_{jl}^{\scriptscriptstyle{U}}(\omega) \langle\langle (\tilde{f}_{j\sigma'}^{\dagger}\tilde{f}_{j\sigma'}-\frac{1}{2})\delta n_{j\overline{\sigma\sigma'}}\tilde{f}_{j\sigma}; \tilde{f}_{i\sigma}^{\dagger}\rangle\rangle_{\omega} \\
        &+ \langle \psi_{l\sigma'}^{\dagger}\tilde{f}_{j\sigma'}\rangle \big(\lambda_{ji} +U\sum_{r\neq j} \lambda_{rj}\langle\langle \psi_{r\sigma}; \tilde{f}_{i\sigma}^{\dagger}\rangle\rangle_{\omega}\big) \\
        &- \sum_{r\neq j}t_{rj}^{\scriptscriptstyle{U}}(\omega)\langle \tilde{f}_{l\sigma'}^{\dagger}\psi_{r\sigma'}\rangle \langle\langle\tilde{f}_{j\sigma}; \tilde{f}_{i\sigma}^{\dagger}\rangle\rangle_{\omega}. 
    \end{split}
\end{equation}
In the first lines of the above two equations, we have made the substitution $\langle\langle\tilde{f}_{j\sigma}; \tilde{f}_{i\sigma}^{\dagger} \rangle\rangle_{\omega} \approx   4\omega\langle\langle\psi_{j\sigma}; \tilde{f}_{i\sigma}^{\dagger} \rangle\rangle_{\omega}/U$, using the Hubbard-I approximation for the trion equation of motion, Eq.~\eqref{eq:eom_psi_freq}.

The following simplifications streamline the expressions. First, we will show in the next section that the second and third lines of Eq.~\eqref{eq:S2} can be combined,
\begin{equation}\label{eq:resonance_broadening_higherorder}
    \begin{split}
        &\frac{U^2}{2\omega}\langle\langle (\tilde{f}_{r \sigma'}^{\dagger}\tilde{f}_{j \sigma'}-\langle\tilde{f}_{r \sigma'}^{\dagger}\tilde{f}_{j \sigma'}\rangle)\delta n_{j \overline{\sigma\sigma'}}\tilde{f}_{j \sigma}; \tilde{f}_{i \sigma}^{\dagger}\rangle\rangle_{\omega}\\
    &\ -2U \langle\langle \delta n_{r \overline{\sigma'}}\tilde{f}_{r \sigma'}^{\dagger}\delta n_{j \overline{\sigma\sigma'}}\tilde{f}_{j \sigma'}\tilde{f}_{j \sigma}; \tilde{f}_{i \sigma}^{\dagger}\rangle\rangle_{\omega}\\
    =& \frac{\omega}{U}\tilde{t}_{rj}^{\scriptscriptstyle U}\langle\langle\psi_{j \sigma}; \tilde{f}_{i \sigma}^{\dagger}\rangle\rangle_{\omega},\quad\quad r\neq j. 
    \end{split}
\end{equation}
Second, by subtracting Eqs.~\eqref{eq:eom_fpsif_freq_N} and~\eqref{eq:eom_psiff_freq_N}, we find that the last three lines of $S^{(1)}(\omega)$ and $S^{(2)}(\omega)$ cancel. The resulting linear equation for $\langle\langle \delta n_{l \overline{\sigma'}}\big(\tilde{f}_{j \sigma'}^{\dagger}\tilde{f}_{l \sigma'}-\tilde{f}_{l \sigma'}^{\dagger}\tilde{f}_{j \sigma'}\big)\tilde{f}_{j \sigma}; \tilde{f}_{i \sigma}^{\dagger}\rangle\rangle_{\omega}$ resembles Eq.~\eqref{eq:eom_nf_freq} and can be solved following the analysis in Appendix~\ref{sec:W}
The solution is a generalization of Eq.~\eqref{eq:gf_psiff},
\begin{equation}\label{eq:resonance_broadening_N}
    \begin{split}
        &\langle\langle \delta n_{l \overline{\sigma'}}\big(\tilde{f}_{j \sigma'}^{\dagger}\tilde{f}_{l \sigma'}-\tilde{f}_{l \sigma'}^{\dagger}\tilde{f}_{j \sigma'}\big)\tilde{f}_{j \sigma}; \tilde{f}_{i \sigma}^{\dagger}\rangle\rangle_{\omega} \\
    \approx&\ \frac{\omega}{U}\sum_{r\neq j}W_{lr, j}(\omega)\left(2t_{rj}^{\scriptscriptstyle{U}}(\omega) + \tilde{t}_{rj}^{\scriptscriptstyle{U}}(\omega)\right)  \langle\langle\psi_{j \sigma}; \tilde{f}_{i \sigma}^{\dagger}\rangle\rangle_{\omega}. \\
    \end{split}
\end{equation}
The resonance broadening then reads
\begin{align}
\langle\langle [\delta n_{j\bar{\sigma}},\tilde{H}]\tilde{f}_{l \sigma}; \tilde{f}_{i  \sigma}^{\dagger}\rangle\rangle= (N-1)c_{r}(\omega)\omega\langle\langle \psi_{j \sigma}; \tilde{f}_{i \sigma}^{\dagger}\rangle\rangle_{\omega},
\end{align}
with the factor $(N-1)$ arising from the flavor summation in Eq.~\eqref{eq:resonance_broadening} and the flavor-independent coefficient
\begin{equation}\label{eq:cr}
    c_{r}(\omega) = \frac{\sum_{\bm k}g(\bm k, \omega) \big[2\bar{\lambda}(\bm k)+\bar{\tilde{\lambda}}(\bm k,\omega)\big]}{\sum_{\bm k}g(\bm k,\omega)}.
\end{equation}
According to the definition below Eq.~\eqref{eq:c1_approx}, $$\bar{\lambda}(\bm k)=\frac{4\omega}{U^2}\sum_{i\neq j}t_{ij}^{\scriptscriptstyle{U}}(\omega)e^{-i\bm k\cdot(\bm R_i-\bm R_j)}. $$ 
Likewise, we define $$\bar{\tilde{\lambda}}(\bm k, \omega) = \frac{4\omega}{U^2}\sum_{i\neq j}\tilde{t}_{ij}^{\scriptscriptstyle{U}}(\omega)e^{-i\bm k\cdot(\bm R_i-\bm R_j)}.$$

\subsection[t]{Evaluation of $\tilde{t}_{rj}^{\scriptscriptstyle{U}}(\omega)$}
In the above analysis of the equations of motion, we have encountered Green's functions in the form of
$$\langle\langle \delta n_{l \overline{\sigma'}}^{\alpha}\tilde{f}_{l \sigma'}^{\dagger}\tilde{f}_{j \sigma'}\delta n_{j \overline{\sigma \sigma'}}^{\beta}\tilde{f}_{j \sigma}; \tilde{f}_{i \sigma}^{\dagger}\rangle\rangle$$
Importantly, it suffices to consider only $\alpha=0,1$ and $\beta<3$ by leveraging the truncations,
\begin{equation}\label{eq:truncation_2}
\langle\langle\delta n_{l \bar{\sigma}}^2...;\tilde{f}_{i\sigma}^{\dagger}\rangle\rangle \approx \frac{1}{4}\langle\langle...;\tilde{f}_{i\sigma}^{\dagger} \rangle\rangle,
\end{equation}
\begin{equation}\label{eq:truncation_3}
\langle\langle\delta n_{j \overline{\sigma\sigma'}}^3...;\tilde{f}_{i\sigma}^{\dagger}\rangle\rangle \approx \langle\langle\delta n_{j \overline{\sigma\sigma'}}...;\tilde{f}_{i\sigma}^{\dagger} \rangle\rangle,
\end{equation}
where the ellipses represent operators that commute with $\delta n_{i\bar{\sigma}}$ [Eq.~\eqref{eq:truncation_2}] or $\delta n_{l\overline{\sigma\sigma'}}$ [Eq.~\eqref{eq:truncation_3}] to the zeroth order in $s^2$. Equation~\eqref{eq:truncation_2} has been used in the main text, e.g., Eq.~\eqref{eq:factorization}. The second approximation is exact for $N=2,4$ and introduces at most $O(s^2)$ errors for $N>4$.

In the calculations, we find it more convenient to define the following Green's functions in the frequency space
\begin{subequations}\label{eq:F}
\begin{align}
    F_{lj,i}^{0,1}(\omega) &=  \langle\langle \tilde{f}_{l \sigma'}^{\dagger}\tilde{f}_{j \sigma'}\delta n_{j \overline{\sigma \sigma'}} \tilde{f}_{j \sigma}; \tilde{f}_{i \sigma}^{\dagger}\rangle\rangle_{\omega}^{c} \notag\\
    &\qquad - \langle\tilde{f}_{l \sigma'}^{\dagger}\tilde{f}_{j \sigma'}\rangle \langle\langle\delta n_{j \overline{\sigma \sigma'}} \tilde{f}_{j \sigma}; \tilde{f}_{i \sigma}^{\dagger}\rangle\rangle_{\omega} ,\label{eq:F01}\\
    F_{lj,i}^{0,2}(\omega)  &=  \langle\langle \tilde{f}_{l \sigma'}^{\dagger}\tilde{f}_{j \sigma'}\delta n_{j \overline{\sigma \sigma'}}^2 \tilde{f}_{j \sigma}; \tilde{f}_{i \sigma}^{\dagger}\rangle\rangle_{\omega}^{c} ,\label{eq:F02}\\
    F_{lj,i}^{1,1}(\omega)  &=  \langle\langle \delta n_{l \overline{\sigma'}}\tilde{f}_{l \sigma'}^{\dagger}\tilde{f}_{j \sigma'}\delta n_{j \overline{\sigma \sigma'}} \tilde{f}_{j \sigma}; \tilde{f}_{i \sigma}^{\dagger}\rangle\rangle_{\omega}^{c}\notag\\
    &\qquad - \langle\psi_{l \sigma'}^{\dagger}\tilde{f}_{j \sigma'}\rangle \langle\langle\delta n_{j \overline{\sigma \sigma'}} \tilde{f}_{j \sigma}; \tilde{f}_{i \sigma}^{\dagger}\rangle\rangle_{\omega}, \label{eq:F11}\\
    F_{lj,i}^{1,2}(\omega)  &=  \langle\langle \delta n_{l \overline{\sigma'}}\tilde{f}_{l \sigma'}^{\dagger}\tilde{f}_{j \sigma'}\delta n_{j \overline{\sigma \sigma'}}^2 \tilde{f}_{j \sigma};\tilde{f}_{i \sigma}^{\dagger}\rangle\rangle_{\omega}^{c} . \label{eq:F12}
\end{align}
\end{subequations}
The definition of the connected Green's function $\langle\langle...\rangle\rangle^{c}$ can be found in Eq.~\eqref{eq:gf_connected}.
To simplify the notation, we omitted the flavor indices in $F^{\alpha,\beta}$ because for the thermally disordered Mott states, its value should be the same $\forall\ (\sigma', \sigma)$ with $\sigma'\neq \sigma$. 
Their equations of motion are derived in Appendix~\ref{sec:eom}. The results are summarized below,
\begin{subequations}\label{eq:eom_F}
\begin{align}
    &\omega F_{lj,i}^{1,1} = -\frac{U}{4}F_{lj,i}^{0,1} + 2UF_{lj,i}^{1,2} + \omega a_{lj}^{11} \langle\langle \psi_{j \sigma}; \tilde{f}_{i \sigma}^{\dagger} \rangle\rangle_{\omega},\label{eq:eom_F11}\\
    &\omega F_{lj,i}^{1,2} = -\frac{U}{4}F_{lj,i}^{0,2} + 2UF_{lj,i}^{1,1} + U a_{lj}^{12} \langle\langle \psi_{j \sigma}; \tilde{f}_{i \sigma}^{\dagger} \rangle\rangle_{\omega},\label{eq:eom_F12}\\
    &\omega F_{lj,i}^{0,1} = -U\sum_{r\neq j} \lambda_{rl} F_{rj,i}^{1,1} + 2U F_{lj,i}^{0,2} + U a_{lj}^{01} \langle\langle \psi_{j \sigma}; \tilde{f}_{i \sigma}^{\dagger}\rangle\rangle_{\omega},\label{eq:eom_F01}\\
    &\omega F_{lj,i}^{0,2} = -U\sum_{r\neq j} \lambda_{rl} F_{rj,i}^{1,2} + 2U F_{lj,i}^{0,1} + \omega a_{lj}^{02}\langle\langle\psi_{j \sigma}; \tilde{f}_{i \sigma}^{\dagger}\rangle\rangle_{\omega},\label{eq:eom_F02}
\end{align}
\end{subequations}
with
\begin{subequations}
\begin{align}
    a_{lj}^{11} &= -\frac{1}{2}\frac{N-2}{N-1}\xi_{lj},\\
    a_{lj}^{12} &= -\frac{3}{8}\frac{N-2}{N-1}\xi_{lj},\\
    a_{lj}^{01} 
    &=\frac{1}{2}\frac{N-2}{N-1}(\lambda_{lj} + \xi_{lj}), \\
    a_{lj}^{02} 
    &= \frac{N-2}{N-1}(\lambda_{lj} - \frac{3}{2}\xi_{lj}).
\end{align}
\end{subequations}
The derivation of these results is quite cumbersome, so we defer the details to a separate section, Appendix~\ref{sec:eom}. In the remainder of the section, we focus on the solutions to these equations.

To solve these equations, we first eliminate $F^{0,j}$ from the last two equations using the first two equations and then apply the method in Appendix~\ref{sec:W} to express 
$$F_{j,i}^{\alpha,\beta}(\bm k,\omega)\equiv \sum_{l\neq j} F_{l j,i}^{\alpha,\beta}(\omega)e^{-i\bm k\cdot(\bm R_l-\bm R_j)}$$ 
in terms of $\langle\langle\psi_{j\sigma }; \tilde{f}_{i\sigma}^{\dagger}\rangle\rangle_{\omega}$. 
The results read as follows:
\begin{subequations}\label{eq:F_solution}
\begin{align}
    F_{j,i}^{0,1}(\bm k,\omega) &= -\frac{4\omega}{U} F_{j,i}^{1,1}(\bm k,\omega) + 8  F_{j,i}^{1,2}(\bm k,\omega) \notag\\
    &\qquad\qquad\qquad + \frac{4\omega}{U} \bar{a}_{\bm k}^{11} \langle\langle\psi_{j\sigma }; \tilde{f}_{i\sigma}^{\dagger}\rangle\rangle_{\omega},\\
    F_{j,i}^{0,2}(\bm k,\omega) &=  -\frac{4\omega}{U}  F_{j,i}^{1,2}(\bm k,\omega) + 8 F_{j,i}^{1,1}(\bm k,\omega) \notag\\
    &\qquad\qquad\qquad + 4\bar{a}_{\bm k}^{12} \langle\langle\psi_{j\sigma }; \tilde{f}_{i\sigma}^{\dagger}\rangle\rangle_{\omega},\\
    F_{j,i}^{1,1}(\bm k,\omega) 
    &=(X_{+}(\bm k, \omega)+ X_{-}(\bm k, \omega))\langle\langle \psi_{j\sigma }; \tilde{f}_{i\sigma}^{\dagger}\rangle\rangle_{\omega},\\
    F_{j,i}^{1,2}(\bm k,\omega) &=(Y_{+}(\bm k, \omega)+ Y_{-}(\bm k, \omega))\langle\langle \psi_{j\sigma }; \tilde{f}_{i\sigma}^{\dagger}\rangle\rangle_{\omega}.
\end{align}
\end{subequations}
with $\bar{a}_{\bm k}^{\alpha\beta} = \sum_{i\neq j} a_{ij}^{\alpha\beta} e^{-i\bm k\cdot(\bm R_i-\bm R_j)}$ and
\begin{widetext}
\begin{subequations}
\begin{align}
    &X_{+}(\bm k, \omega) = \frac{U}{2}\bigg[\frac{1}{(\omega-2U)^2-\frac{U^2\lambda(\bm k)}{4}} + \frac{1}{(\omega+2U)^2-\frac{U^2\lambda(\bm k)}{4}}\bigg]\left[\frac{\omega^2}{U} \bar{a}_{\bm k}^{11}-U(2\bar{a}_{\bm k}^{12}+\frac{1}{4}\bar{a}_{\bm k}^{01})-\bar{X}_{+}(\omega)\right],\\
    &X_{-}(\bm k, \omega) = \frac{U}{2}\bigg[\frac{1}{(\omega-2U)^2-\frac{U^2\lambda(\bm k)}{4}} - \frac{1}{(\omega+2U)^2-\frac{U^2\lambda(\bm k)}{4}}\bigg]\left[\omega \bar{a}_{\bm k}^{12}-\omega (2\bar{a}_{\bm k}^{11}+\frac{1}{4}\bar{a}_{\bm k}^{02})-\bar{X}_{-}(\omega)\right],\\
    &Y_{+}(\bm k, \omega) = \frac{U}{2}\bigg[\frac{1}{(\omega-2U)^2-\frac{U^2\lambda(\bm k)}{4}} + \frac{1}{(\omega+2U)^2-\frac{U^2\lambda(\bm k)}{4}}\bigg]\left[\omega \bar{a}_{\bm k}^{12}-\omega (2\bar{a}_{\bm k}^{11}+\frac{1}{4}\bar{a}_{\bm k}^{02})-\bar{Y}_{+}(\omega)\right],\\
    &Y_{-}(\bm k, \omega) = \frac{U}{2}\bigg[\frac{1}{(\omega-2U)^2-\frac{U^2\lambda(\bm k)}{4}} - \frac{1}{(\omega+2U)^2-\frac{U^2\lambda(\bm k)}{4}}\bigg]\left[\frac{\omega^2}{U} \bar{a}_{\bm k}^{11}-U(2\bar{a}_{\bm k}^{12}+\frac{1}{4}\bar{a}_{\bm k}^{01})-\bar{Y}_{-}(\omega)\right].
\end{align}
\end{subequations} 
\end{widetext}
Here $\bar{X}_{\pm}(\omega)$ and $\bar{Y}_{\pm}(\omega)$ are determined by conditions $$\sum_{\bm k}X_{\pm}(\bm k,\omega) = \sum_{\bm k}Y_{\pm}(\bm k,\omega) = 0.$$

Based on Eqs.~\eqref{eq:F} and~\eqref{eq:F_solution}, we can now derive Eq.~\eqref{eq:resonance_broadening_higherorder}.
\begin{equation}
    \frac{U^2}{2\omega}F_{rj,i}^{0,1}-2UF_{rj,i}^{1,1} = \frac{\omega}{U} \tilde{t}_{rj}^{\scriptscriptstyle{U}}(\omega)\langle\langle \psi_{j\sigma }; \tilde{f}_{i\sigma}^{\dagger}\rangle\rangle_{\omega},
\end{equation}
where $\tilde{t}_{rj}^{\scriptscriptstyle{U}}(\omega) = \frac{U^2}{4\omega}\int_{\bm k}\bar{\tilde{\lambda}}(\bm k,\omega) e^{i\bm k\cdot (\bm R_r - \bm R_j)}$ and
\begin{equation}\label{eq:bartildelambda}
\begin{split}
    \bar{\tilde{\lambda}}(\bm k,\omega) =&-\frac{64\omega}{U}(X_{+}(\bm k, \omega)+X_{-}(\bm k, \omega))\\
    &+32(2Y_{+}(\bm k, \omega) + 2Y_{-}(\bm k, \omega)+\bar{a}_{\bm k}^{11}).
\end{split}
\end{equation}
It is straightforward to verify that $$\tilde{t}_{jj}^{\scriptscriptstyle{U}} \propto \sum_{\bm k}\bar{\tilde{\lambda}}(\bm k,\omega)=0.$$ 

Note that $X(\bm k,\omega),Y(\bm k,\omega), \bar{\tilde{\lambda}}(\bm k,\omega)$ are real for $|\omega|<U/2$.
Therefore, to the first order in $s^2$, the imaginary part of $c_r(\omega)$ obeys
\begin{subequations}
\begin{align}
    c_r^{\prime\prime}(\omega) &\approx \frac{\sum_{\bm k}g(\bm k,\omega)^{\prime\prime}(2\bar{\lambda}(\bm k)+\bar{\tilde{\lambda}}(\bm k,\omega))}{\sum_{\bm k}g(\bm k,\omega)^{\prime}} \notag\\
    &=-\pi s^2 \text{sgn}(\omega)\Theta\Big(\frac{U^2}{4}-\omega^2\Big) \frac{\frac{U^2}{4}}{\frac{U^2}{4}-\omega^2}\notag\\
    &\qquad\quad\quad\ \times (2\bar{\lambda}(\bm k)+\bar{\tilde{\lambda}}(\bm k,\omega))|_{\hat{\lambda}(\bm k)=\frac{4\omega^2}{U^2}}.\label{eq:c2_im}
\end{align}
\end{subequations}
%
%
%
%
Plugging Eq.~\eqref{eq:bartildelambda} into the above formula and after a lengthy calculation, 
we find a remarkably simple relation for $|\omega|<U/2$,
\begin{equation}\label{eq:c_r_im}
    c_{r}''(\omega) = \frac{N}{N-1} c''(\omega) + O(s^2),
\end{equation}
where $c(\omega)$ is given in Eq.~\eqref{eq:c1_approx}. Furthermore, we neglect the imaginary part of $c_{r}(\omega)$ at higher energies, $|\omega|>U/2$, which are unimportant for quasiparticle lifetime. Within this approximation, Eq.~\eqref{eq:c_r_im} further implies $c_{r}(\omega) \approx [N/(N-1)]c(\omega)$. Plugging this relation into Eq.~\eqref{eq:cr}, we obtain the resonance broadening correction Eq.~\eqref{eq:resonance_broadening_correction} for general even $N$. 

\section[Equations of motion]{Equations of motion for $F^{\alpha,\beta}$}\label{sec:eom}
\subsection[F11]{$F^{1,1}$}
In this section, we derive the equation of motion for $F_{lj,i}^{1,1}$. First, in the definition Eq.~\eqref{eq:F11}, the connected Green's function is actually identical to the full Green's function, because $\langle \delta n_{l \overline{\sigma'}}\tilde{f}_{l \sigma'}^{\dagger}\tilde{f}_{j \sigma'}\delta n_{j \overline{\sigma \sigma'}}\rangle = 0$. The equation of motion for the first term in Eq.~\eqref{eq:F11} then reads
\begin{align}\label{eq:eom_F11_approx}
        & i\partial_t \langle\langle \delta n_{l \overline{\sigma'}}\tilde{f}_{l \sigma'}^{\dagger}\tilde{f}_{j \sigma'}\delta n_{j \overline{\sigma \sigma'}}\tilde{f}_{j\sigma}; \tilde{f}_{i \sigma}^{\dagger}\rangle\rangle \notag\\
        %
        %
        \approx & -U\langle\langle \delta n_{l\overline{\sigma'}}^2\tilde{f}_{l \sigma'}^{\dagger}\delta n_{j \overline{\sigma \sigma'}}\tilde{f}_{j \sigma'}\tilde{f}_{j\sigma}; \tilde{f}_{i \sigma}^{\dagger}\rangle\rangle\notag\\
        &+U\langle\langle \delta n_{l\overline{\sigma'}}\tilde{f}_{l \sigma'}^{\dagger}\tilde{f}_{j \sigma'}\delta n_{j \overline{\sigma \sigma'}}\delta n_{j \bar{\sigma}}\tilde{f}_{j\sigma}; \tilde{f}_{i \sigma}^{\dagger}\rangle\rangle\notag\\
        &+ U\langle\langle \delta n_{l\overline{\sigma'}}\tilde{f}_{l \sigma'}^{\dagger}\delta n_{j \overline{\sigma'}}\tilde{f}_{j \sigma'}\delta n_{j \overline{\sigma \sigma'}}\tilde{f}_{j\sigma}; \tilde{f}_{i \sigma}^{\dagger}\rangle\rangle\notag\\
        &+U\lambda_{jl}\langle\langle \delta n_{l \overline{\sigma'}}^2f_{l\sigma'}^{\dagger}f_{l\sigma'} \delta n_{j \overline{\sigma \sigma'}}\tilde{f}_{j\sigma}; \tilde{f}_{i \sigma}^{\dagger}\rangle\rangle.
\end{align}
On the right-hand side of the equation, we retain only Green's functions that contain fermionic operators on at most two different sites. This means that (i) we drop the commutators $[\delta n_{l\overline{\sigma'}}, \tilde{H}]$ and $[\delta n_{j\overline{\sigma\sigma'}}, \tilde{H}]$ in the equation of motion; (ii) For commutators $[\tilde{f}_{l\sigma'}^{\dagger},\tilde{H}]$  and $[\tilde{f}_{j\sigma},\tilde{H}]$, we neglect the terms that change the sites in Eq.~\eqref{eq:commutator_fH}, and the remaining terms become the first two terms on the right-hand side of the equation, respectively; (iii) For commutators $[\tilde{f}_{j\sigma'},\tilde{H}]$, we need to keep both the onsite scattering term and the term that transfers electrons at site $j$ to site $l$. They correspond to the last two terms on the right-hand side. All the dropped terms give rise to corrections of higher order in $s^2$.

The right-hand side of Eq.~\eqref{eq:eom_F11_approx} can be further simplified by combining the second and third terms using
\begin{equation}
    \tilde{f}_{j \sigma'}\delta n_{j \bar{\sigma}}\tilde{f}_{j\sigma} + \delta n_{j \overline{\sigma'}}\tilde{f}_{j \sigma'}\tilde{f}_{j\sigma} = 2\delta n_{j\overline{\sigma\sigma'}}\tilde{f}_{j \sigma'}\tilde{f}_{j\sigma}.
\end{equation}
After some algebra, we arrive at
\begin{equation}\label{eq:eom_F11_approx2}
\begin{split}
    &i\partial_t \langle\langle \delta n_{l \overline{\sigma'}}\tilde{f}_{l \sigma'}^{\dagger}\tilde{f}_{j \sigma'}\delta n_{j \overline{\sigma \sigma'}}\tilde{f}_{j\sigma}; \tilde{f}_{i \sigma}^{\dagger}\rangle\rangle\\
    \approx& -\frac{U}{4}\langle\langle \tilde{f}_{l \sigma'}^{\dagger}\tilde{f}_{j \sigma'}\delta n_{j \overline{\sigma \sigma'}}\tilde{f}_{j\sigma}; \tilde{f}_{i \sigma}^{\dagger}\rangle\rangle  \\
    &+ 2U\langle\langle \delta n_{l \overline{\sigma'}}\tilde{f}_{l \sigma'}^{\dagger}\tilde{f}_{j \sigma'}\delta n_{j\overline{\sigma \sigma'}}^2\tilde{f}_{j\sigma}; \tilde{f}_{i \sigma}^{\dagger}\rangle\rangle\\
    & + \frac{U}{8}\lambda_{lj}\langle\langle \delta n_{j \overline{\sigma \sigma'}}\tilde{f}_{j\sigma}; \tilde{f}_{i \sigma}^{\dagger}\rangle\rangle.
\end{split}
\end{equation}
The last term arises from factorizing the Green's function in the last line of Eq.~\eqref{eq:eom_F11_approx},
\begin{equation}
\begin{split}
    &\langle\langle \delta n_{l \overline{\sigma'}}^2f_{l\sigma'}^{\dagger}f_{l\sigma'} \delta n_{j \overline{\sigma \sigma'}}\tilde{f}_{j\sigma}; \tilde{f}_{i \sigma}^{\dagger}\rangle\rangle\\
    \approx&\ \langle \delta n_{l \overline{\sigma'}}^2\rangle\langle f_{l\sigma'}^{\dagger}f_{l\sigma'} \rangle\langle\langle\delta n_{j \overline{\sigma \sigma'}}\tilde{f}_{j\sigma}; \tilde{f}_{i \sigma}^{\dagger}\rangle\rangle.
\end{split}
\end{equation}
Because the flavor moment in the Mott state is thermally disorder, the following relation holds
\begin{align}\label{eq:nf_psi_conversion}
    \langle\langle \delta n_{j \overline{\sigma \sigma'}} \tilde{f}_{j\sigma}; \tilde{f}_{i \sigma}^{\dagger}\rangle\rangle &= \frac{N-2}{N-1}\langle\langle \psi_{j\sigma}; \tilde{f}_{i \sigma}^{\dagger}\rangle\rangle.
\end{align}
After Fourier transforming the above three equations to the frequency space and expressing the Green's functions in Eqs.~\eqref{eq:F} in terms of $F^{\alpha,\beta}$ defined in Eq.~\eqref{eq:F}, we arrive at the equation of motion of $F^{1,1}$, Eq.~\eqref{eq:eom_F11}.
%

\subsection[F12]{$F^{1,2}$}
The equation of motion for $F^{1,2}$ are derived using a procedure similar to that for $F^{1,1}$:
\begin{align}
    &i\partial_t \langle\langle \big(\delta n_{l \overline{\sigma'}}\tilde{f}_{l \sigma'}^{\dagger}\tilde{f}_{j \sigma'}\delta n_{j \overline{\sigma \sigma'}}^2\tilde{f}_{j\sigma}; \tilde{f}_{i \sigma}^{\dagger}\rangle\rangle^c \notag\\
    %
    %
    \approx &-\frac{U}{4}\langle\langle \tilde{f}_{r \sigma'}^{\dagger}\tilde{f}_{j \sigma'}\delta n_{j \overline{\sigma \sigma'}}^2\tilde{f}_{j\sigma}; \tilde{f}_{i \sigma}^{\dagger}\rangle\rangle \notag\\
    &+ 2U \langle\langle \delta n_{l\overline{\sigma'}}\tilde{f}_{l \sigma'}^{\dagger}\tilde{f}_{j \sigma'}\delta n_{j \overline{\sigma \sigma'}}^3\tilde{f}_{j\sigma}; \tilde{f}_{i \sigma}^{\dagger}\rangle\rangle \notag\\
    & - U \langle \delta n_{l \overline{\sigma'}}\tilde{f}_{l \sigma'}^{\dagger}\tilde{f}_{j \sigma'}\delta n_{j \overline{\sigma \sigma'}}^2\rangle \langle\langle \delta n_{j\bar{\sigma}}\tilde{f}_{j\sigma},\tilde{f}_{i\sigma}^{\dagger}\rangle\rangle \notag\\
    &+ U\lambda_{jl}\langle\langle \delta n_{l \overline{\sigma'}}^2f_{l\sigma'}^{\dagger}f_{l\sigma'} \delta n_{j \overline{\sigma \sigma'}}^2 \tilde{f}_{j\sigma}; \tilde{f}_{i \sigma}^{\dagger}\rangle\rangle \notag\\
    &- U \sum_{r\neq l}\lambda_{rl}\langle\langle \delta n_{l\overline{\sigma'}}\delta n_{r\overline{\sigma'}}\tilde{f}_{r \sigma'}^{\dagger}\tilde{f}_{j \sigma'}\delta n_{j \overline{\sigma \sigma'}}^2\tilde{f}_{j\sigma}; \tilde{f}_{i \sigma}^{\dagger}\rangle\rangle\notag\\
    &+ U \sum_{r\neq j,l}\lambda_{jr}\langle\langle \delta n_{l\overline{\sigma'}}\tilde{f}_{l \sigma'}^{\dagger}\delta n_{r\overline{\sigma'}}\tilde{f}_{r \sigma'}\delta n_{j \overline{\sigma \sigma'}}^2\tilde{f}_{j\sigma}; \tilde{f}_{i \sigma}^{\dagger}\rangle\rangle\notag\\
    &+ U \sum_{r\neq j}\lambda_{jr}\langle\langle \delta n_{l\overline{\sigma'}}\tilde{f}_{l \sigma'}^{\dagger}\tilde{f}_{j \sigma'}\delta n_{j \overline{\sigma \sigma'}}^2\delta n_{r \bar{\sigma}}\tilde{f}_{r\sigma}; \tilde{f}_{i \sigma}^{\dagger}\rangle\rangle\notag\\
    & - U \langle \delta n_{l \overline{\sigma'}}\tilde{f}_{l \sigma'}^{\dagger}\tilde{f}_{j \sigma'}\delta n_{j \overline{\sigma \sigma'}}^2\rangle  \sum_{r\neq j}\lambda_{jr}\langle\langle \delta n_{r\bar{\sigma}}\tilde{f}_{r\sigma},\tilde{f}_{i\sigma}^{\dagger}\rangle\rangle.\label{eq:eom_F12_approx}
\end{align}
We kept only the terms generated by the commutators $[\tilde{f}_{l\sigma'}^{\dagger},\tilde{H}], [\tilde{f}_{j\sigma'},\tilde{H}]$, and $[\tilde{f}_{j\sigma},\tilde{H}]$. The summations in the last four lines exclude terms at $r=j$ or $l$. Those excluded terms are combined into the first four lines on the right-hand side of the equation. The fifth and sixth lines on the right-hand side generically involve three different sites and are of higher order in $s^2$. Moreover, the last two lines cancel to the desired accuracy due to the factorization for $r\neq j,l$,
\begin{equation}
    \begin{split}
        &\langle\langle \delta n_{l\overline{\sigma'}}\tilde{f}_{l \sigma'}^{\dagger}\tilde{f}_{j \sigma'}\delta n_{j \overline{\sigma \sigma'}}^2\delta n_{r \bar{\sigma}}\tilde{f}_{r\sigma}; \tilde{f}_{i \sigma}^{\dagger}\rangle\rangle\\
        \approx &\ \langle \delta n_{l\overline{\sigma'}}\tilde{f}_{l \sigma'}^{\dagger}\tilde{f}_{j \sigma'}\delta n_{j \overline{\sigma \sigma'}}^2\rangle \langle\langle\delta n_{r \bar{\sigma}}\tilde{f}_{r\sigma}; \tilde{f}_{i \sigma}^{\dagger}\rangle\rangle.
    \end{split}
\end{equation}
Next, we simplify the first four lines on the right-hand side of Eq.~\eqref{eq:eom_F12_approx} by truncating the correlation functions using Eq.~\eqref{eq:truncation_3} and the following factorizations:
\begin{equation}
\begin{split}
    &\langle\langle \delta n_{l \overline{\sigma'}}^2f_{l\sigma'}^{\dagger}f_{l\sigma'} \delta n_{j \overline{\sigma \sigma'}}^2 \tilde{f}_{j\sigma}; \tilde{f}_{i \sigma}^{\dagger}\rangle\rangle \\
    \approx&\ \langle \delta n_{l \overline{\sigma'}}^2\rangle \langle f_{l\sigma'}^{\dagger}f_{l\sigma'} \rangle \langle\delta n_{j \overline{\sigma \sigma'}}^2 \rangle\langle\langle\tilde{f}_{j\sigma}; \tilde{f}_{i \sigma}^{\dagger}\rangle\rangle,
\end{split}
\end{equation}
and
\begin{equation}\label{eq:decoupling_psifn2}
    \langle \delta n_{l \overline{\sigma'}}\tilde{f}_{l \sigma'}^{\dagger}\tilde{f}_{j \sigma'}\delta n_{j \overline{\sigma \sigma'}}^{2}\rangle \approx \langle \delta n_{l \overline{\sigma'}}\tilde{f}_{l \sigma'}^{\dagger}\tilde{f}_{j \sigma'}\rangle\langle\delta n_{j \overline{\sigma \sigma'}}^{2}\rangle,
\end{equation}
with
\begin{equation}\label{eq:n_N-2_squared}
    \langle\delta n_{j \overline{\sigma \sigma'}}^{2}\rangle = \frac{1}{2}\frac{N-2}{N-1}.
\end{equation}
Eq.~\eqref{eq:eom_F12_approx} reduces to
\begin{align}
    &i\partial_t \langle\langle \big(\delta n_{l \overline{\sigma'}}\tilde{f}_{l \sigma'}^{\dagger}\tilde{f}_{j \sigma'}\delta n_{j \overline{\sigma \sigma'}}^2\tilde{f}_{j\sigma}; \tilde{f}_{i \sigma}^{\dagger}\rangle\rangle^c \notag\\
    \approx &-\frac{U}{4}\langle\langle \tilde{f}_{r \sigma'}^{\dagger}\tilde{f}_{j \sigma'}\delta n_{j \overline{\sigma \sigma'}}^2\tilde{f}_{j\sigma}; \tilde{f}_{i \sigma}^{\dagger}\rangle\rangle \notag\\
    &+ 2U \langle\langle \delta n_{l\overline{\sigma'}}\tilde{f}_{l \sigma'}^{\dagger}\tilde{f}_{j \sigma'}\delta n_{j \overline{\sigma \sigma'}}\tilde{f}_{j\sigma}; \tilde{f}_{i \sigma}^{\dagger}\rangle\rangle \notag\\
    & - U \langle \psi_{l \sigma'}^{\dagger}\tilde{f}_{j \sigma'}\rangle\langle\delta n_{j \overline{\sigma \sigma'}}^2\rangle \langle\langle \delta n_{j\bar{\sigma}}\tilde{f}_{j\sigma},\tilde{f}_{i\sigma}^{\dagger}\rangle\rangle \notag\\
    &+ \frac{U}{2}\lambda_{jl}\langle \delta n_{l \overline{\sigma'}}^2\rangle\langle\delta n_{j \overline{\sigma \sigma'}}^2\rangle \langle\langle\tilde{f}_{j\sigma}; \tilde{f}_{i \sigma}^{\dagger}\rangle\rangle.\label{eq:eom_F12_approx2}
\end{align}
We then Fourier transform Eq.~\eqref{eq:eom_F12_approx2} to frequency space and, using the definitions in Eq.~\eqref{eq:F12}, recast it as the equation of motion for $F^{1,2}$, Eq.~\eqref{eq:eom_F12}.

\subsection[F01]{$F^{0,1}$}
Here, we compute the equation of motion for $F^{0,1}$. First,
\begin{align}
    &i\partial_t \langle\langle \tilde{f}_{l \sigma'}^{\dagger}\tilde{f}_{j \sigma'}\delta n_{j \overline{\sigma \sigma'}} \tilde{f}_{j\sigma}; \tilde{f}_{i \sigma}^{\dagger}\rangle\rangle^{c} \notag\\
    %
    \approx
    &\ 2U \langle\langle \tilde{f}_{l \sigma'}^{\dagger}\tilde{f}_{j\sigma'}\delta n_{j \overline{\sigma\sigma'}}^2\tilde{f}_{j\sigma}; \tilde{f}_{i \sigma}^{\dagger}\rangle\rangle \notag\\
    &- U \langle\tilde{f}_{l \sigma'}^{\dagger}\tilde{f}_{j \sigma'}\delta n_{j \overline{\sigma \sigma'}}\rangle \langle\langle \psi_{j\sigma}; \tilde{f}_{i \sigma}^{\dagger}\rangle\rangle \notag\\
    &-U\lambda_{jl}\langle\langle \delta n_{j \overline{\sigma'}} \tilde{f}_{j \sigma'}^{\dagger}\tilde{f}_{j\sigma'}\delta n_{j \overline{\sigma\sigma'}}\tilde{f}_{j\sigma}; \tilde{f}_{i \sigma}^{\dagger}\rangle\rangle\notag\\
    &- U\sum_{r\neq j}\lambda_{rl}\langle\langle \delta n_{r \overline{\sigma'}} \tilde{f}_{r \sigma'}^{\dagger}\tilde{f}_{j\sigma'}\delta n_{j \overline{\sigma\sigma'}}\tilde{f}_{j\sigma}; \tilde{f}_{i \sigma}^{\dagger}\rangle\rangle \notag\\
    &+ U\sum_{r\neq j} \lambda_{jr}  \langle\langle\tilde{f}_{l \sigma'}^{\dagger}\delta n_{r \overline{\sigma'}}\tilde{f}_{r \sigma'} \delta n_{j \overline{\sigma \sigma'}}\tilde{f}_{j\sigma}; \tilde{f}_{i \sigma}^{\dagger}\rangle\rangle.
    \label{eq:eom_F01_approx}
\end{align}
We have separated the terms at $r=j$ from the summations. The equation of motion generates higher-order Green's functions on the right-hand side of Eq.~\eqref{eq:eom_F01_approx}, which we analyze in turn.

The first term can be expressed in terms of $F^{0,2}(\omega)$ using the relation
\begin{equation}\label{eq:g_ffnnf}
\begin{split}
    &\langle\langle \tilde{f}_{l \sigma'}^{\dagger}\tilde{f}_{j\sigma'}\delta n_{j \overline{\sigma\sigma'}}^2\tilde{f}_{j\sigma}; \tilde{f}_{i \sigma}^{\dagger}\rangle\rangle_{\omega} \\
    =&\ F_{lj,i}^{0,2}(\omega)+ \langle \tilde{f}_{l \sigma'}^{\dagger}\tilde{f}_{j\sigma'}\rangle \langle\delta n_{j \overline{\sigma\sigma'}}^2\rangle \langle\langle\tilde{f}_{j\sigma}; \tilde{f}_{i \sigma}^{\dagger}\rangle\rangle_{\omega},
\end{split}
\end{equation}
where we applied Eq.~\eqref{eq:n_N-2_squared}.
%

The third term on the right-hand side of Eq.~\eqref{eq:eom_F01_approx} can be rewritten as follows:
\begin{align}\label{eq:g_psipsif}
    &\langle\langle \delta n_{j \overline{\sigma'}} \tilde{f}_{j \sigma'}^{\dagger}\tilde{f}_{j\sigma'}\delta n_{j \overline{\sigma\sigma'}}\tilde{f}_{j\sigma}; \tilde{f}_{i \sigma}^{\dagger}\rangle\rangle \notag\\
    =& \langle\langle \delta n_{j \overline{\sigma\sigma'}}^2 \delta n_{j\sigma'}\tilde{f}_{j\sigma}; \tilde{f}_{i \sigma}^{\dagger}\rangle\rangle +\frac{1}{2}\langle\langle \delta n_{j \overline{\sigma\sigma'}}^2 \tilde{f}_{j\sigma}; \tilde{f}_{i \sigma}^{\dagger}\rangle\rangle  \notag\\
    &-\frac{1}{2}\langle\langle \delta n_{j \overline{\sigma\sigma'}}\delta n_{j\sigma'}\tilde{f}_{j\sigma}; \tilde{f}_{i \sigma}^{\dagger}\rangle\rangle - \frac{1}{4}\langle\langle \delta n_{j \overline{\sigma\sigma'}} \tilde{f}_{j\sigma}; \tilde{f}_{i \sigma}^{\dagger}\rangle\rangle,
\end{align}
with $\delta n_{i\sigma} = f_{i\sigma}^{\dagger}f_{i\sigma}-1/2$. Because
\begin{align}
    \delta n_{i\bar{\sigma}}^2&=(\delta n_{i\overline{\sigma\sigma'}}+ \delta n_{i\sigma})^2=\frac{1}{4}+\delta n_{i\overline{\sigma\sigma'}}^2 + 2 \delta n_{i\overline{\sigma\sigma'}}\delta n_{i\sigma},
\end{align}
and the substitution $\delta n_{i\bar{\sigma}}^2\rightarrow 1/4$ in the Green's functions is accurate up to $O(s^2)$ corrections, 
we can make the following substitution in Eq.~\eqref{eq:g_psipsif},
\begin{equation}\label{eq:trunction_2}
    \langle\langle\delta n_{i\overline{\sigma\sigma'}}\delta n_{i\sigma}...,\tilde{f}_{i\sigma}^{\dagger}\rangle\rangle\approx -\frac{1}{2}\langle\langle\delta n_{i\overline{\sigma\sigma'}}^2...,\tilde{f}_{i\sigma}^{\dagger}\rangle\rangle.
\end{equation}
Together with the approximation Eq.~\eqref{eq:truncation_3}, we find that
\begin{align}\label{eq:g_psipsif2}
    &\langle\langle \delta n_{j \overline{\sigma'}} \tilde{f}_{j \sigma'}^{\dagger}\tilde{f}_{j\sigma'}\delta n_{j \overline{\sigma\sigma'}}\tilde{f}_{j\sigma}; \tilde{f}_{i \sigma}^{\dagger}\rangle\rangle \notag\\
    \approx&\ \frac{3}{4}\Big(\langle \delta n_{j\overline{\sigma\sigma'}}^2\rangle\langle\langle\tilde{f}_{j\sigma}; \tilde{f}_{i \sigma}^{\dagger}\rangle\rangle -  \frac{N-2}{N-1}\langle\langle\psi_{j\sigma}; \tilde{f}_{i \sigma}^{\dagger}\rangle\rangle \Big).
\end{align}
Here we used Eq.~\eqref{eq:nf_psi_conversion} for flavor-disordered Mott states.

The fourth term on the right-hand side of Eq.~\eqref{eq:eom_F01_approx} can be rewritten in terms of $F^{1,1}$ via the relation
\begin{align}\label{eq:eom_F01_fifth}
    &\sum_{r\neq j}\lambda_{rl}\langle\langle\delta n_{r \overline{\sigma'}} \tilde{f}_{r \sigma'}^{\dagger}\tilde{f}_{j\sigma'}\delta n_{j \overline{\sigma\sigma'}}\tilde{f}_{j\sigma}; \tilde{f}_{i \sigma}^{\dagger}\rangle\rangle_{\omega} \notag\\
     =& \sum_{r\neq j}\lambda_{rl} F_{rj,i}^{1,1}(\omega) + \frac{N-2}{N-1}\sum_{r\neq j}\lambda_{rl}\langle \psi_{r \sigma'}^{\dagger}\tilde{f}_{j\sigma'}\rangle\langle\langle\psi_{j\sigma}; \tilde{f}_{i \sigma}^{\dagger}\rangle\rangle_{\omega}.
\end{align}

In the last line of Eq.~\eqref{eq:eom_F01_approx}, we factorize the Green's function and obtain
\begin{equation}\label{eq:eom_F01_sixth}
\begin{split}
    &\sum_{r\neq j}\lambda_{jr}\langle\langle \tilde{f}_{l \sigma'}^{\dagger}\delta n_{r \overline{\sigma'}}\tilde{f}_{r\sigma'}\delta n_{j \overline{\sigma\sigma'}}\tilde{f}_{j\sigma}; \tilde{f}_{i \sigma}^{\dagger}\rangle\rangle_{\omega} \\
    \approx&\ \frac{N-2}{N-1}\sum_{r\neq j}\lambda_{jr}\langle\tilde{f}_{l \sigma'}^{\dagger}\psi_{r\sigma'}\rangle \langle\langle \psi_{j\sigma}; \tilde{f}_{i \sigma}^{\dagger}\rangle\rangle_{\omega}.
\end{split}
\end{equation}
%

Next, we plug Eqs.~\eqref{eq:g_ffnnf},~\eqref{eq:trunction_2},~\eqref{eq:eom_F01_fifth} and~\eqref{eq:fifth_sixth} into the frequency-space Fourier transform of Eq.~\eqref{eq:eom_F01_approx}.
In the resulting expression, the last term of Eq.~\eqref{eq:eom_F01_fifth} and the right-hand side of Eq.~\eqref{eq:eom_F01_sixth} can be combined via the following relation,
\begin{equation}\label{eq:fifth_sixth}
    \sum_{r\neq j}\lambda_{jr}\langle\tilde{f}_{l \sigma'}^{\dagger}\psi_{r\sigma'}\rangle - \sum_{r\neq j}\lambda_{rl}\langle \psi_{r \sigma'}^{\dagger}\tilde{f}_{j\sigma'}\rangle \approx \frac{1}{4}(\xi_{jl} - \lambda_{jl}),
\end{equation}
where we used $\langle \psi_{j \sigma'}^{\dagger}\tilde{f}_{r\sigma'}\rangle = \langle \tilde{f}_{j\sigma'}^{\dagger}\psi_{r \sigma'}\rangle = -\xi_{rj}/4$ [see Eq.~\eqref{eq:corr_fpsi}] and the approximation $\xi_{jj}=1+O(s^2)$. Furthermore, we can make the substitution $\langle\langle\tilde{f}_{j\sigma}; \tilde{f}_{i\sigma}^{\dagger} \rangle\rangle_{\omega} \approx   4\omega\langle\langle\psi_{j\sigma}; \tilde{f}_{i\sigma}^{\dagger} \rangle\rangle_{\omega}/U$. After some algebra, we find that Eq.~\eqref{eq:eom_F01_approx} becomes
\begin{equation}\label{eq:eom_F01_approx2}
\begin{split}
    &\omega\langle\langle \tilde{f}_{l \sigma'}^{\dagger}\tilde{f}_{j \sigma'}\delta n_{j \overline{\sigma \sigma'}} \tilde{f}_{j\sigma}; \tilde{f}_{i \sigma}^{\dagger}\rangle\rangle_{\omega}^{c}\\
    =&-U\sum_{r\neq j} \lambda_{rl} F_{r j,i}^{1,1}(\omega) + 2U F_{lj,i}^{0,2}(\omega) \\
    &+ \frac{N-2}{N-1}\big[\frac{\lambda_{lj} + \xi_{lj}}{2}U-\frac{\xi_{lj}}{4}\omega\big]\langle\langle \psi_{j\sigma}; \tilde{f}_{i \sigma}^{\dagger}\rangle\rangle_{\omega}.
\end{split}
\end{equation}
Finally, using the definition of $F^{0,1}$ Eq.~\eqref{eq:F02}, we obtain the equation of motion for $F^{0,1}$, Eq.~\eqref{eq:eom_F01}, from Eq.~\eqref{eq:eom_F01_approx2}.
%
%

\subsection[F02]{$F^{0,2}$}
The derivations of equation of motion for $F^{0,2}$ are analgous to that for $F^{0,1}$:
\begin{align}
    &i\partial_t \langle\langle \tilde{f}_{l \sigma'}^{\dagger}\tilde{f}_{j \sigma'}\delta n_{j \overline{\sigma \sigma'}}^2\tilde{f}_{j\sigma}; \tilde{f}_{i \sigma}^{\dagger}\rangle\rangle^{c} \notag\\
    \approx& 
    2U \langle\langle \tilde{f}_{l \sigma'}^{\dagger}\tilde{f}_{j\sigma'}\delta n_{j \overline{\sigma\sigma'}}^3\tilde{f}_{j\sigma}; \tilde{f}_{i \sigma}^{\dagger}\rangle\rangle \notag\\
    &- U \langle\tilde{f}_{l \sigma'}^{\dagger}\tilde{f}_{j \sigma'}\rangle  \langle\langle \delta n_{j \overline{\sigma \sigma'}}^2\delta n_{j \bar{\sigma}}\tilde{f}_{j\sigma}; \tilde{f}_{i \sigma}^{\dagger}\rangle\rangle\notag\\
    &-U\lambda_{jl}\langle\langle \delta n_{j \overline{\sigma'}} \tilde{f}_{j \sigma'}^{\dagger}\tilde{f}_{j\sigma'}\delta n_{j \overline{\sigma\sigma'}}^2\tilde{f}_{j\sigma}; \tilde{f}_{i \sigma}^{\dagger}\rangle\rangle\notag\\
    &-U\sum_{r\neq j}\lambda_{rl}\langle\langle \delta n_{r \overline{\sigma'}} \tilde{f}_{r \sigma'}^{\dagger}\tilde{f}_{j\sigma'}\delta n_{j \overline{\sigma\sigma'}}^2\tilde{f}_{j\sigma}; \tilde{f}_{i \sigma}^{\dagger}\rangle\rangle \notag\\
    &+ U\sum_{r\neq j} \lambda_{jr}  \langle\langle\tilde{f}_{l \sigma'}^{\dagger}\tilde{f}_{r \sigma'}\delta n_{r \overline{\sigma'}} \delta n_{j \overline{\sigma \sigma'}}^2\tilde{f}_{j\sigma}; \tilde{f}_{i \sigma}^{\dagger}\rangle\rangle.\label{eq:eom_F02_approx}
\end{align}
We now analyze the higher-order Green's function on the right-hand side.

The first term can be reduced to a lower-order Green's function using the truncation Eq.~\eqref{eq:truncation_3}. The result can be rewritten in terms of $F^{1,1}$ as follows,
\begin{equation}\label{eq:g_ffnnnf}
\begin{split}
    &\langle\langle \tilde{f}_{l \sigma'}^{\dagger}\tilde{f}_{j\sigma'}\delta n_{j \overline{\sigma\sigma'}}^3\tilde{f}_{j\sigma}; \tilde{f}_{i \sigma}^{\dagger}\rangle\rangle_{\omega}\\ 
    =&\ F_{lj,i}^{0,1}(\omega)+\langle\tilde{f}_{l \sigma'}^{\dagger}\psi_{j \sigma'}\rangle \langle\langle\tilde{f}_{j \sigma}; \tilde{f}_{i \sigma}^{\dagger}\rangle\rangle_{\omega}\\
    &+ \frac{N-2}{N-1}\langle\tilde{f}_{l \sigma'}^{\dagger}\tilde{f}_{j \sigma'}\rangle \langle\langle\psi_{j \sigma}; \tilde{f}_{i \sigma}^{\dagger}\rangle\rangle_{\omega}.
\end{split}
\end{equation}

The third line of Eq.~\eqref{eq:eom_F02_approx} can be rewritten as follows:
\begin{align}\label{eq:g_psipsinf2}
    &\langle\langle \delta n_{j \overline{\sigma'}} \tilde{f}_{j \sigma'}^{\dagger}\tilde{f}_{j\sigma'}\delta n_{j \overline{\sigma\sigma'}}^2\tilde{f}_{j\sigma}; \tilde{f}_{i \sigma}^{\dagger}\rangle\rangle \notag\\
    \approx&\ \frac{3}{4}\Big(\frac{N-2}{N-1}\langle\langle\psi_{j\sigma}; \tilde{f}_{i \sigma}^{\dagger}\rangle\rangle - \langle \delta n_{j\overline{\sigma\sigma'}}^2\rangle\langle\langle\tilde{f}_{j\sigma}; \tilde{f}_{i \sigma}^{\dagger}\rangle\rangle \Big).
\end{align}

The fourth term can be rewritten in terms of $F^{1,2}$,
\begin{align}
    &\sum_{r\neq j}\lambda_{rl}\langle\langle\delta n_{r \overline{\sigma'}} \tilde{f}_{r \sigma'}^{\dagger}\tilde{f}_{j\sigma'}\delta n_{j \overline{\sigma\sigma'}}^2\tilde{f}_{j\sigma}; \tilde{f}_{i \sigma}^{\dagger}\rangle\rangle_{\omega} \notag\\
    =& \sum_{r\neq j}\lambda_{rl} F_{rj,i}^{1,2} + \sum_{r\neq j}\lambda_{jr} \langle \tilde{f}_{l\sigma'}^{\dagger}\psi_{r \sigma'}\rangle \langle \delta n_{j \overline{\sigma\sigma'}}^2\rangle \langle\langle\tilde{f}_{j\sigma}; \tilde{f}_{i \sigma}^{\dagger}\rangle\rangle_{\omega}. \label{eq:g_psipsinf_sum}
\end{align}

In the last line of Eq.~\eqref{eq:eom_F02_approx}, we factorize the Green's function and obtain
\begin{align}\label{eq:eom_F02_sixth}
    &\sum_{r\neq j} \lambda_{jr}  \langle\langle\tilde{f}_{l \sigma'}^{\dagger}\tilde{f}_{r \sigma'}\delta n_{r \overline{\sigma'}} \delta n_{j \overline{\sigma \sigma'}}^2\tilde{f}_{j\sigma}; \tilde{f}_{i \sigma}^{\dagger}\rangle\rangle \notag\\
    =&\sum_{r\neq j} \lambda_{jr}  \langle\tilde{f}_{l \sigma'}^{\dagger}\psi_{r \sigma'}\rangle \langle\delta n_{j \overline{\sigma \sigma'}}^2\rangle \langle\langle \tilde{f}_{j\sigma}; \tilde{f}_{i \sigma}^{\dagger}\rangle\rangle_{\omega}
\end{align}

Plugging Eqs.~\eqref{eq:g_ffnnnf}-~\eqref{eq:eom_F02_sixth} and using the definition Eq.~\eqref{eq:F02}, we arrive at the equation of motion for $F^{0,2}$, Eq.~\eqref{eq:eom_F02}
%

\bibliographystyle{apsrev4-2}
\bibliography{reference}
\end{document}